\documentclass{aa}
\usepackage[utf8]{inputenc}

\title{The host of the Type I SLSN 2017egm: A young, sub-solar metallicity environment in a massive spiral galaxy.}
\author{L. Izzo\textsuperscript{1}, C. C. Th\"one\textsuperscript{1}, R. Garc\'ia-Benito\textsuperscript{1}, A. de Ugarte Postigo\textsuperscript{1,2}, Z. Cano\textsuperscript{1}, D. A. Kann\textsuperscript{1}, K. Bensch \textsuperscript{1},  M. Della Valle\textsuperscript{3,4}, D.~Galad\'i-Enr\'iquez\textsuperscript{5}, R. P. Hedrosa\textsuperscript{5}}
\titlerunning{The host galaxy of SLSN 2017egm}
\authorrunning{L. Izzo et al.}

\institute{\textsuperscript{1}Instituto de Astrofisica de Andalucia (IAA-CSIC), Glorieta de la Astronomia s/n, 18008 Granada, Spain \\
\textsuperscript{2}Dark Cosmology Centre, University of Copenhagen, Juliane Maries Vej 30, 2100 Copenhagen \O, Denmark\\
\textsuperscript{3}INAF-Capodimonte Observatory, Salita Moiariello 16, I-80131, Napoli, Italy\\
\textsuperscript{4}International Center for Relativistic Astrophysics, Piazzale della Repubblica, 2, 65122, Pescara, Italy\\
\textsuperscript{5}Centro Astron\'omico Hispano-Alem\'an, Calar Alto, CSIC-MPG, Sierra de los Filabres-04550 G\'ergal (Almer\'ia), Spain}

\abstract{Type-I Super-luminous-Supernova (SLSN) host galaxies are predominantly low-metallicity, highly star-forming dwarfs. One of the current key questions is whether SLSNe Type I can only occur in such environments and hosts.}
{Here we present an integral-field study of the massive, high-metallicity spiral NGC 3191, the host of SN 2017egm, the closest SLSN Type I to date. We use data from PMAS/CAHA and the public MaNGA survey to shed light on the properties of the SLSN site and the origin of star-formation in this non-starburst spiral galaxy.
}
{We map the physical properties different \ion{H}{II} regions throughout the galaxy and characterize their  stellar populations using the STARLIGHT fitting code. Kinematical information allows to study a possible interaction with its neighbouring galaxy as the origin of recent star formation activity which could have caused the SLSN. 
}
{NGC 3191 shows intense star-formation in the western part with three large SF regions of low metallicity. Taking only the properties of emitting gas, the central regions of the host have a higher metallicity, lower specific star-formation rate and lower ionization. Modeling the stellar populations gives a different picture: The SLSN region has two dominant stellar populations with different ages, the youngest one with an age of 2-10 Myr and lower metallicity, likely the population from which the SN progenitor originated. Emission line kinematics of NGC 3191 show indications of interaction with its neighbour MCG+08-19-017 at $\sim$45 kpc, which might be responsible for the recent starburst. In fact, this galaxy pair has in total hosted 4 SNe, 1988B (Type Ia), SN 2003ds (Type Ic in MCG+08-19-017), PTF10bgl (SLSN-Type II) and 2017egm, underlying the enhanced SF in both galaxies due to interaction.
}{Our study shows that one has to be careful interpreting global host and even gas properties without looking at the stellar population history of the region. SLSNe seem to still be consistent with massive stars ($>$ 20 M$_\odot$) requiring low ($< 0.6Z_{\odot}$) metallicity and those environments can also occur in massive, late-type galaxies but not necessarily starbursts.}

\begin{document}


\date{}

\maketitle

\section{Introduction}
Superluminous supernovae (SLSNe) are defined as supernovae (SNe) with absolute peak magnitudes in any optical band of $<-21$ mag \citep{GalYam2012} and were detected for the first time a decade ago thanks to the advent of new untargeted large-scale surveys probing all types of galaxies \citep{Kaiser2005,Drake2009,Law2009,Tonry2012}. SLSNe have been divided into three main classes: 1) SLSNe-Type I or H-poor SNe characterized by the absence of hydrogen in the spectra \citep{Quimby2007,Quimby2011,Pastorello2010ApJ,Chomiuk2011}; 2) SLSNe-Type II or H-rich that show narrow hydrogen lines, likely due to a massive surrounding H-envelope, therefore basically making them high-luminosity Type IIn SNe \citep{Ofek2007,Smith2007}, and possibly 3) SLSN-Type ``R'', SLSNe powered by radioactive decay of large amounts of Ni. Their high luminosities imply that their origins are different from those of normal SNe, and the study of the immediate environment of these events can provide important information on physical properties of their progenitors which so far have remained undetected.

Indeed, several studies of SLSN-host-galaxy samples have revealed significant differences in the properties of Type-I and Type-II hosts \citep{Leloudas2015,Perley2016,Lunnan2014, Schulze2016}. Type-I hosts are usually star-forming dwarf galaxies with low metallicities at the extreme end of the distribution in the Baldwin-Philips-Terlevich \citep[BPT;][]{Baldwin1981} diagram and likely even more extreme than Gamma-Ray-Burst (GRB) hosts (see, e.g., \citealt{Leloudas2015, Schulze2016} but see also \citealt{Lunnan2014}). In addition, they have been found in very young stellar populations, indicating that SLSNe Type-I might be the first and most massive stars to end in a SN explosion after the onset of a new starburst episode in their galaxy or HII region \citep{Thoene2015}. Type-II hosts seem to be more metal rich and within the bulk of emission-line galaxies in the SDSS. At higher redshifts, SLSN-I hosts evolve in line with the most star-forming galaxies towards higher-mass galaxies with higher absolute star-formation rates, but still seem to show a metallicity cutoff at $\sim0.4$ solar and a tendency towards low-mass hosts \citep{Schulze2016}. 

In this paper we present Integral-Field Unit (IFU) spectroscopic data of NGC 3191, the host galaxy of SN 2017egm, a Type-I SLSN discovered by the Gaia satellite \citep{Delgado2017} on May 23. Detailed analysis of its early optical spectra revealed its SLSN-Type-I nature \citep{DongATEL, Nicholl2017}. The SLSN reached its optical maximum on June 20, 2017 with an absolute magnitude of $M_U \sim -22$ mag. As pointed out in \cite{Nicholl2017}, \cite{Bose2017}, and \cite{Chen2017c}, the host seems to be the first Type I host with super-solar metallicity, on which we will comment in the paper. In this work, we analyse the properties of different regions in the host galaxy, as well as the SLSN location, providing a more detailed analysis of the metallicity and the star-formation rate distributions around the SLSN environment and several other \ion{H}{II} regions. Resolved spectral modeling across the host galaxy gives us important clues on the different underlying stellar populations, their ages and metallicities, allowing us a more detailed insight on this apparently very
peculiar SLSN-I environment.

\section{Observations}

\begin{figure*}[ht]
    \centering
    \includegraphics[width=0.98\columnwidth]{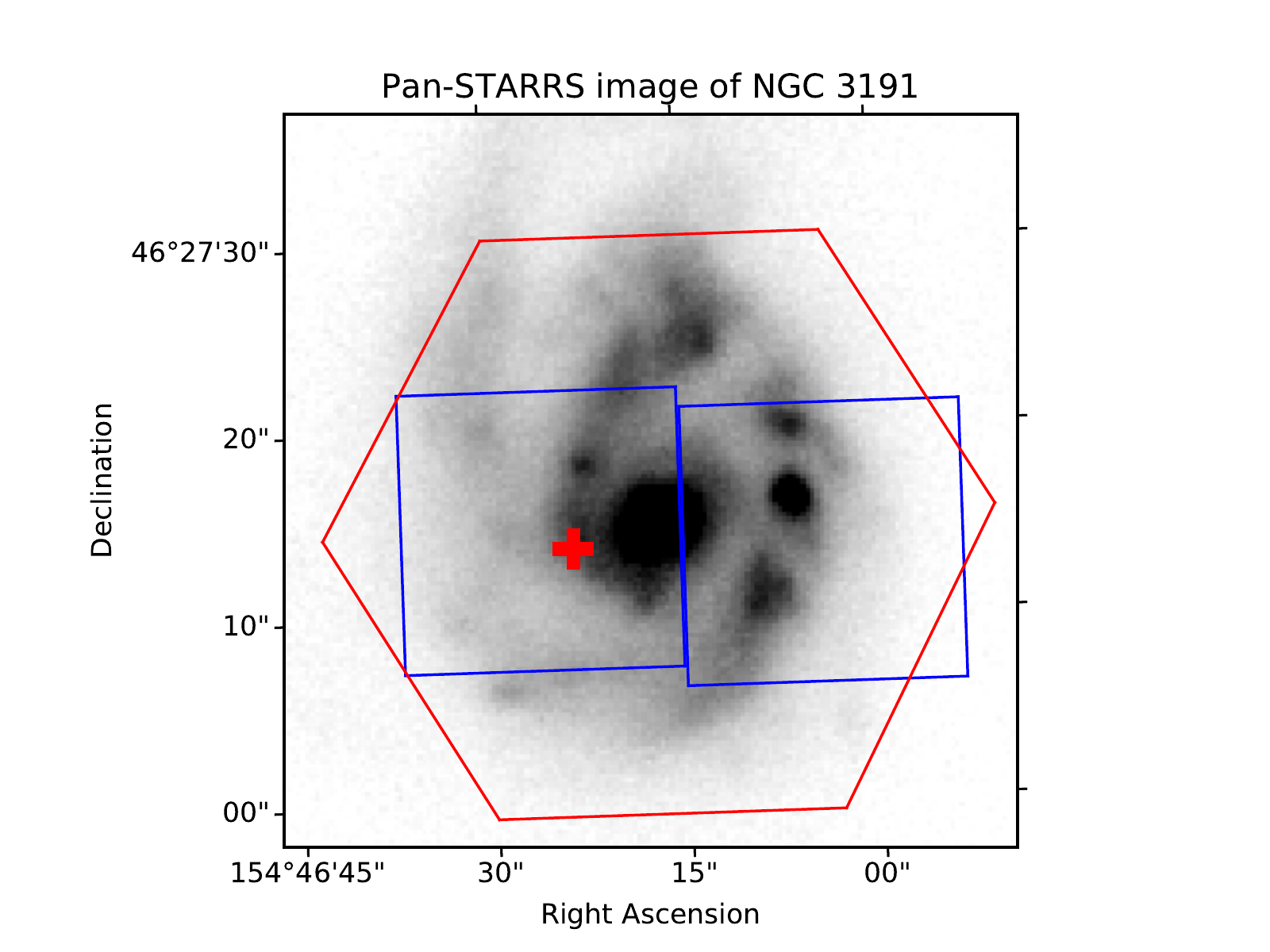}
    \includegraphics[width=\columnwidth]{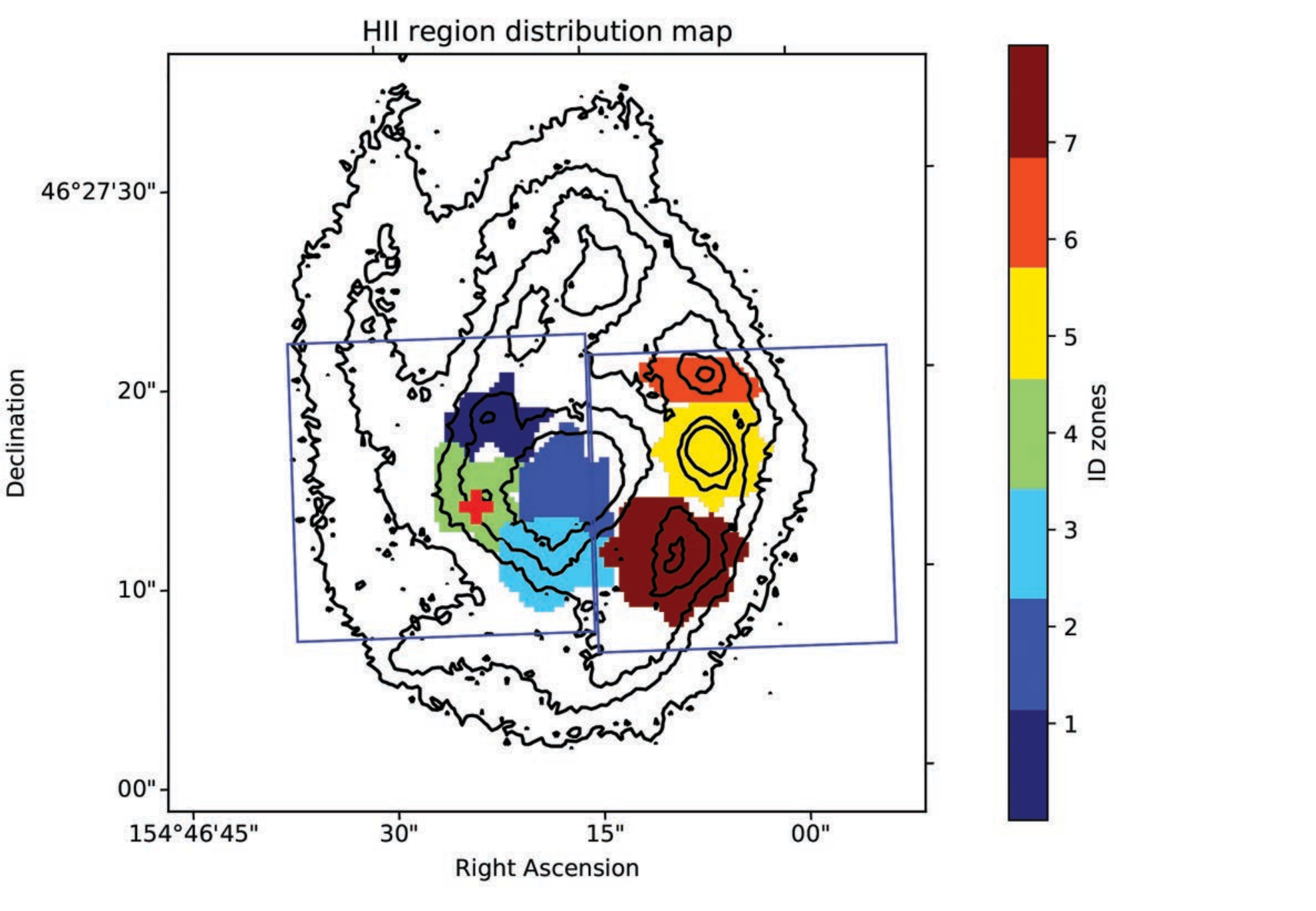}
    \caption{\textit{Left panel}: The Pan-STARRS $r^\prime$-band image of NGC 3191 with the two PMAS fields-of-view (blue squares) and the MaNGA one (red hexagon) superposed. \textit{Right panel}:  Map of the \ion{H}{II} regions as obtained with the code described in the text and applied to the PMAS cubes. The SLSN 2017egm position is marked with a red cross in both panels. The levels of the isophotes, from the outermost to the innermost ones, are 21.86, 21.11, 20.66, 20.22 and 19.91 mag/arcsec, respectively. }
    \label{fig:no1}
\end{figure*}

We observed the host galaxy of SN 2017egm with the Potsdam Multi-Aperture Spectrophotometer (PMAS) mounted on the 3.5m telescope of the Calar Alto Observatory (Spain) \citep{Roth2005} on June 21, 2017. PMAS is an integral field spectrograph composed of $16\times16$ square elements in the lens-array configuration. We used the 1\farcs0 spatial resolution configuration, which provides a field of view of $16^{\prime\prime}\times16^{\prime\prime}$; the R600 grism was positioned at the grating position 146.5, which results in the wavelength range $4700-7700$ \AA\, with a spectral resolution of 1.55 \AA\, (resolving power R $\sim$ 600) and the maximum efficiency centred around the rest-frame H$\alpha$ region ($\sim6600$ \AA). The average seeing during the observations was of 1\farcs2. The data were reduced with the \texttt{P3D} data-reduction tools \citep{Sandin2010} to obtain a final flux-calibrated cube for each single observation. 

Due to the dimensions of the host galaxy NGC 3191 ($46^{\prime\prime}\times37^{\prime\prime}$) we were not able to cover the entire galaxy within a single PMAS shot. Given the limited time available, we decided to observe the SLSN region and the adjacent western half of the galaxy characterized by the presence of active \ion{H}{II} regions, with two single exposures of 600 s each. The galaxy and the two regions covered by our PMAS observations are shown in the left panel of Fig. \ref{fig:no1}. 

We complement our dataset with the publicly available observations of NGC 3191 from the Mapping Nearby Galaxies at APO (MaNGA) survey \citep{Law2015}, available in DR14 of the SDSS\footnote{http://skyserver.sdss.org/dr14/en/tools/explore/Summary. aspx?id=1237658613587181631}. The host galaxy, of which about 2/3 are covered by the MaNGA data, was observed in January 2016, well before the appearance of the SLSN. Given the lower spatial resolution of the MaNGA survey (on average 2\farcs5)
we used the PMAS data for the identification and extraction of the \ion{H}{II} regions in the host. For properties that need a higher spectral resolution (1.0 \AA, R $\sim$ 2000), such as the estimate of the velocity distribution and stellar population properties, we used MaNGA data. To ensure extraction of similar regions in the MaNGA as in the PMAS data we projected the World Coordinate System (WCS) of the Pan-STARRS image on the data cubes of both PMAS and MaNGA using the \texttt{astropy} python package \citep{astropy}, whose set of libraries has been extensively used in this work.

Finally, we use three public spectra of NGC 3191 and its neighbour galaxy MCG+08-19-017. The first spectrum covers the centre of NGC 3191 and has been obtained in the context of the BOSS survey \citep[SDSS-III,][]{Dawson2013}, with 2$^{\prime\prime}$ fibres and an average resolving power of $\textnormal{R}\sim2000$ at $\sim6000$ \AA. The other two spectra cover the brightest SF region in the west of NGC 3191 and the centre of MCG+08-19-017 and have been obtained within the Legacy Survey \citep[SDSS-I/II,][]{York2000}, with 3$^{\prime\prime}$ fibres and an average resolving power of $\textnormal{R}\sim1800$.

\begin{table}[ht]
    \centering
    \begin{tabular}{lcc}
\hline
 Region        & RA             & DEC             \\
\hline
 \ion{H}{II}-1 & 10h 19m 07.90s & +46d 26m 48.35s \\
 \ion{H}{II}-2 & 10h 19m 07.81s & +46d 26m 47.32s  \\
 \ion{H}{II}-3 & 10h 19m 07.84s & +46d 26m 46.58s \\
 \ion{H}{II}-4 & 10h 19m 07.93s & +46d 26m 47.36s \\
 \ion{H}{II}-5 & 10h 19m 08.00s & +46d 26m 48.14s \\
 \ion{H}{II}-6 & 10h 19m 08.00s & +46d 26m 49.14s \\
 \ion{H}{II}-7 & 10h 19m 08.05s & +46d 26m 46.91s \\
\hline
\end{tabular}
    \caption{Coordinates (J2000.0) of the emission peaks for each \ion{H}{II} region that we identified. The \ion{H}{II}-2 region corresponds to the galaxy centre while \ion{H}{II}-4 is the SLSN site.}
    \label{tab:no1}
\end{table}

\section{Emission line analysis}

\begin{figure}
    \centering
    \includegraphics[width=\columnwidth]{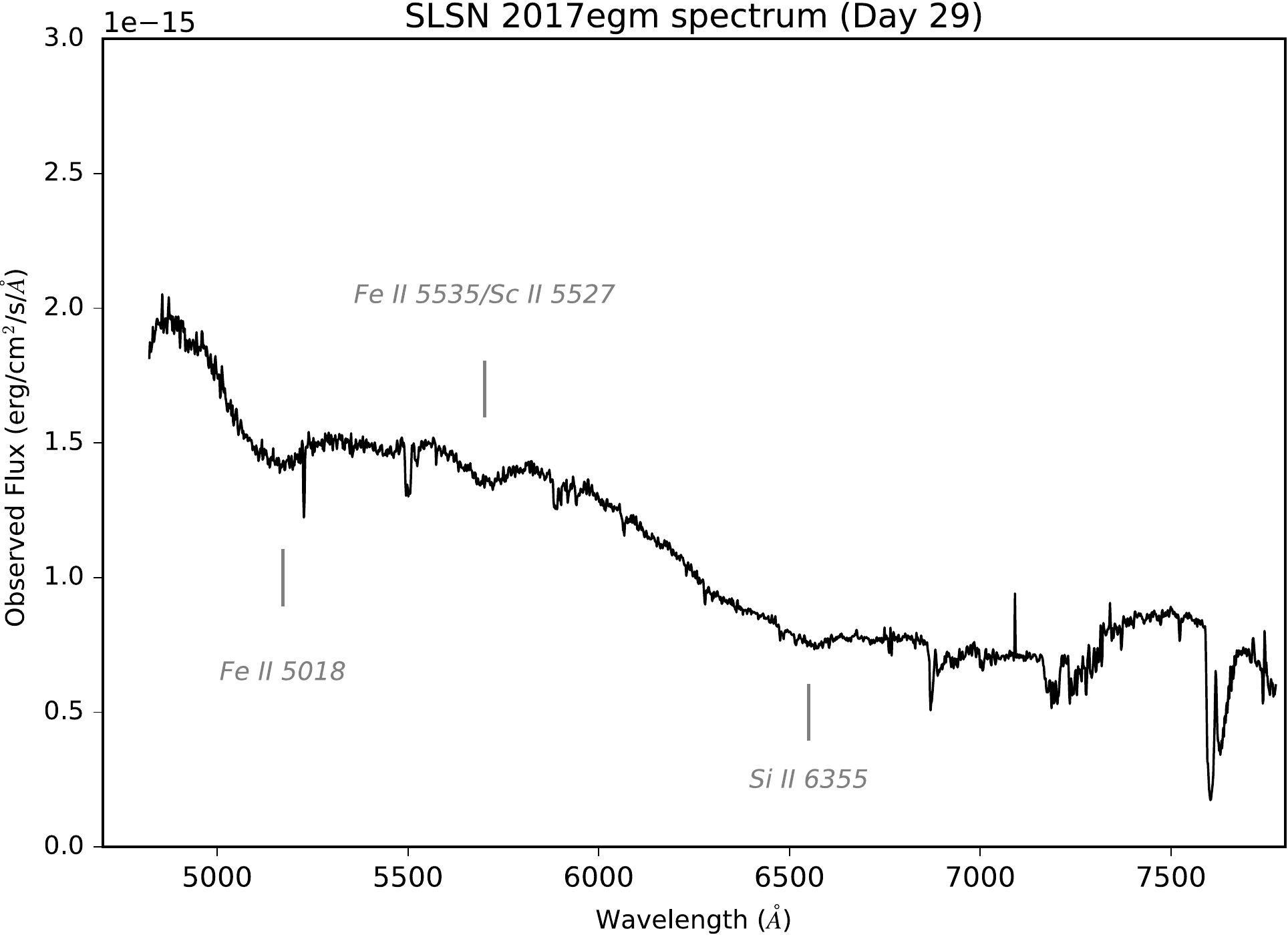}
    \caption{The observed spectrum of SLSN 2017egm 29 days after discovery, which we use as template for correcting the PMAS data of region \ion{H}{II}-4. We identified some typical features of Type-I SLSNe such as \ion{Fe}{II} $5018\lambda$, the blend \ion{Fe}{II} $5535\lambda$ - \ion{Sc}{II} $5527\lambda$ and \ion{Si}{II} $6355\lambda$.}
    \label{fig:no0}
\end{figure}

The host galaxy of SN 2017egm is a spiral galaxy (type Sbc) at a redshift of $z = 0.03072$ \citep{Hakobyan2012} with an absolute magnitude of $M_r = -21.9$ \citep{Taddia2016} and a stellar mass of $M_{*} = 10^{10.61} M_{\odot}$ \citep{Stoll2013}. The Pan-STARRS archival $r^\prime$-band image reveals the presence of several bright knots corresponding to highly star-forming \ion{H}{II} regions in the galaxy. From the PMAS cubes we create a map of the H$\alpha$ distribution by summing the flux of 15 pixels in wavelength centred at the corresponding redshifted emission line. The continuum is subtracted by taking the average flux of 10 pixels red- and bluewards with respect to the H$\alpha$ line and free of sky emission or telluric absorption lines.

In the H$\alpha$ map we identify seven \ion{H}{II} regions using a code derived from the \texttt{HII-explorer} package\footnote{For more details see http://lucagrb.weebly.com/hii.html} taking an H$\alpha$ threshold for the peak flux for each region of $P = 3 \times 10^{-16}$ erg/cm$^2$/s, corresponding to an intrinsic luminosity of 5 $\times$ 10$^{38}$ erg/s (similarly as it was done in \citet{Izzo2017}). The identified \ion{H}{II} regions are shown in the right panel of Fig. \ref{fig:no1}, their central coordinates are reported in Table \ref{tab:no1}. \ion{H}{II} regions 2 and 4 correspond to the galaxy centre and the SLSN position, respectively. Region 4 indeed shows spectral features from the actual SLSN, which, at the time of our observations was close to its maximum luminosity (see Fig. \ref{fig:no0}). We also note that the \ion{H}{II}-6 region has some defects in MaNGA data which affect the extraction of an integrated spectrum of this region.

To correct the PMAS data for the underlying SN emission, we extract the SLSN spectrum from the PMAS data using the following process: We interpolate MaNGA data with a power-law function to fit the continuum of the underlying \ion{H}{II} region. Then we subtract this model from the PMAS \ion{H}{II}-4 region, after a pass by the \texttt{ZAP} software for the subtraction of sky and telluric emission lines \citep{Soto2016}. The final SN spectrum is shown in Fig. \ref{fig:no0}. Finally, in order to accurately measure line fluxes, the SN spectrum has been subtracted from the \ion{H}{II}-4 region.

\begin{figure}
    \centering
    \includegraphics[width=\columnwidth]{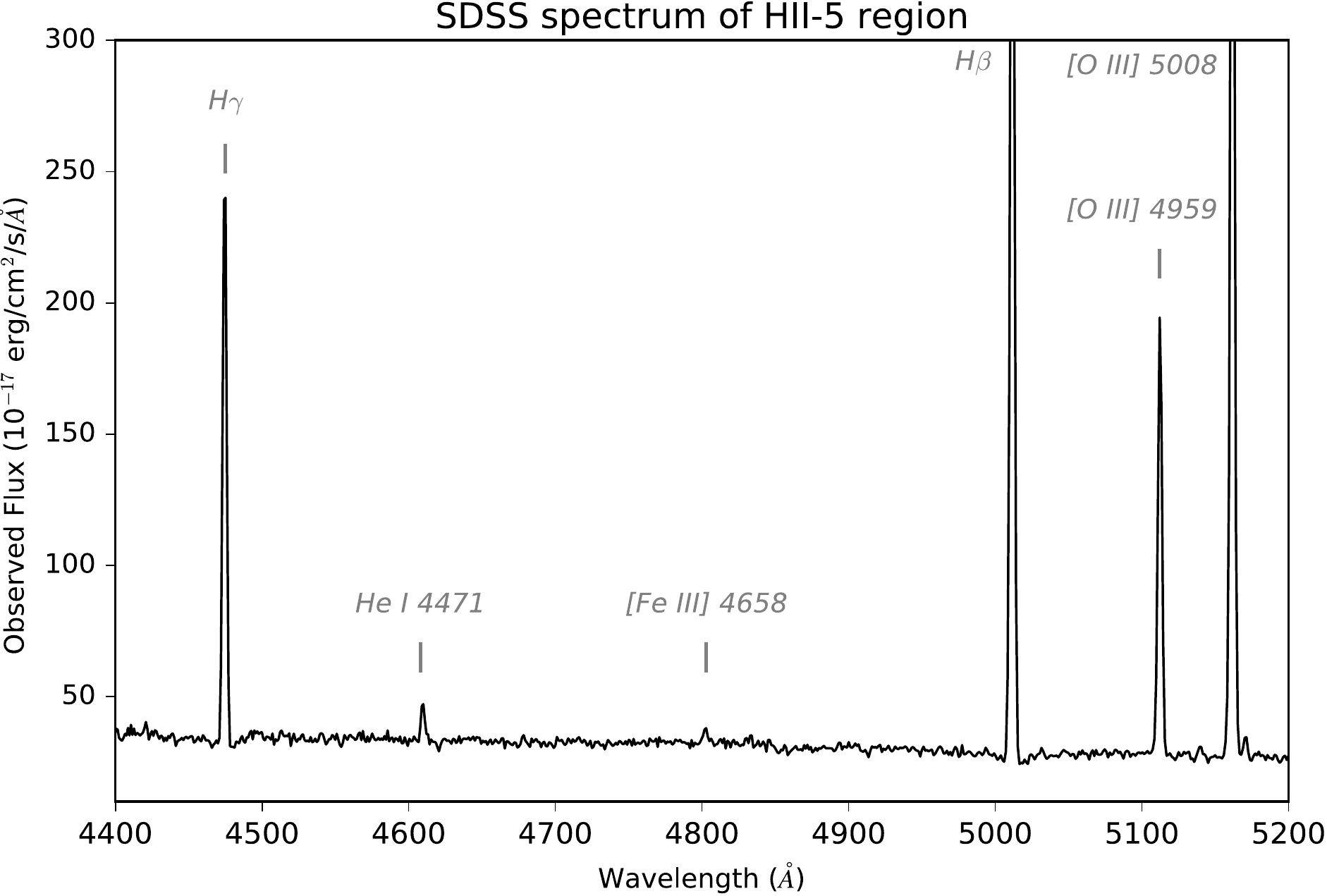}
    \caption{Section of the SDSS spectrum of region \ion{H}{II}-5, showing \ion{He}{I} $4471\lambda$ and a faint absorption trough at the bottom of the bright H$\beta$ line.}
    \label{fig:noSDSS}
\end{figure}

We do not find any absorption feature caused by an underlying stellar population around the H$\beta$ line in the PMAS data cube, even in the spectra with the highest S/N value (\ion{H}{II}-5,6,7 regions). This is mainly caused by the low spectral resolution and by the low signal-to-noise ratio of some of the spectra. A spectrum of the region \ion{H}{II}-5 can also be found in the SDSS-DR12 Science Archive Server\footnote{The spectrum is available at the link: https://dr12.sdss.org/spectrumDetail?mjd=52614 \&fiber=502\&plateid=944}. In this spectrum we note the  presence of a broad absorption feature at the H$\beta$ line with a measured equivalent width (EW) of EW$_{H\beta,*}=3.8$ \AA. We also find a faint \ion{He}{I} $4471\lambda$ line with an EW$_{HeI}=1.3$ \AA{} (see also Fig. \ref{fig:noSDSS}). Similar results are obtained from the MaNGA data of the \ion{H}{II}-5 region. In particular, we detect a faint \ion{He}{I} 4471$\lambda$ line in the spectrum of the SLSN region with EW$_{HeI}=0.7$ \AA{}. Since the presence of helium is an indicator of the young age of the underlying stellar population, we estimate the age following the method underlined in \citet{GonzalezDelgado1999}: this method assumes an instantaneous burst of star-formation (10$^6$ M$_{\odot}$) at solar metallicity, with an initial mass function modelled with a power-law function using a decay rate of $\alpha=-2.35$ and with mass cut-offs at $M_{low}=1M_{\odot}$ and $M_{up}=80M_{\odot}$. With these parameters we obtain an estimate for the stellar age of around 7 Myr; for lower metallicities we obtain similar results. 

For an additional age estimate we use the stellar population synthesis code presented in \citet{Levesque2013} which gives age estimates related to the H$\alpha$ EWs. The EW values for the \ion{H}{II} regions 1--3 are lower by a factor of 5 to 10 (for the galaxy centre) than the EWs measured for the brightest \ion{H}{II} regions 5--7 (see Table \ref{tab:no2}), indicating a younger stellar age of the latter regions. Assuming a metallicity of $Z=0.008$ corresponding to $\sim 0.6Z_{\odot}$, the derived stellar population ages for the \ion{H}{II} regions 5--7 are indeed younger ($\sim6-7$ Myr) than the \ion{H}{II} regions in the rest of the galaxy ($\geq10-20$ Myr). The SLSN region shows an intermediate value, implying that its stellar age is younger than adjacent regions but older than the most active ones. This evidence is also in agreement with the detection of helium at the SLSN site.

Subsequently, we corrected our PMAS \ion{H}{II} spectra for the underlying stellar populations using as templates the stellar models provided by the \texttt{SED}$@$ code \citep{Cervino2004}. These models consider a Salpeter initial mass function in the mass range $0.1-120M_\odot$ and use the Padova isochrone tables \citep{Girardi2002} for an age of 7 Myr for the regions 5, 6 and 7, and an age of 15 Myr for the remaining regions. Given the better spatial resolution of PMAS data, we used the segmentation map derived from PMAS to extract \ion{H}{II} region spectra from MaNGA data. 
We measured the emission line fluxes with the SHerpa IFU line fitting software (SHIFU; García-Benito, in preparation), based on CIAOs Sherpa package \citep{Freeman2001, Doe2007} by fitting single gaussians to residual spectra obtained after subtracting the \texttt{STARLIGHT} fit from the observed spectrum (see description below). A first order polynomial was used to fit the continuum baseline in order to take into account small deviations of the fit. Measuring the emission lines in the residual spectra, particularly Balmer and Paschen lines, corrects their fluxes affected by the presence of absorption wings of stellar origin. The list of the main emission lines identified in the spectra of each region is shown in Table \ref{tab:no2}, while in the Table \ref{tab:App1} we report additional line measurements.

\begin{table*}
    \centering
    \begin{tabular}{lcccccccc}
\hline
 Region        & Hbeta           & [\ion{O}{III}] 5007   & [\ion{N}{II}] 6550   & Halpha          & [\ion{N}{II}] 6586   & [\ion{S}{II}] 6718   & [\ion{S}{II}] 6732  & EW(H$\alpha$)  \\
\hline
 \ion{H}{II}-1 & 1.63$\pm$0.05  & 0.61$\pm$0.04       & 0.80$\pm$0.04      & 6.70$\pm$0.10  & 2.39$\pm$0.06      & 1.13$\pm$0.04      & 0.78$\pm$0.04  &  42.4  \\
 \ion{H}{II}-2 & 2.52$\pm$0.11  & 0.61$\pm$0.10       & 1.46$\pm$0.17      & 11.45$\pm$0.27 & 4.12$\pm$0.16      & 1.72$\pm$0.0.11    & 1.25$\pm$0.14  &  21.0   \\
 \ion{H}{II}-3 & 2.41$\pm$0.08  & 0.58$\pm$0.05       & 1.08$\pm$0.05      & 10.13$\pm$0.25 & 3.29$\pm$0.09      & 1.48$\pm$0.04      & 1.05$\pm$0.06  &  41.9  \\
 \ion{H}{II}-4 & 1.71$\pm$0.05  & 1.24$\pm$0.04       & 0.59$\pm$0.12      & 6.61$\pm$0.15  & 1.87$\pm$0.04      & 1.08$\pm$0.02      & 0.74$\pm$0.02  &  104.9  \\
 \ion{H}{II}-5 & 14.67$\pm$0.06 & 14.83$\pm$0.31      & 4.79$\pm$0.12      & 52.10$\pm$0.80 & 14.02$\pm$0.25     & 7.72$\pm$1.42      & 5.81$\pm$0.13  &  189.4     \\
 \ion{H}{II}-6 & 1.39$\pm$0.03  & 1.38$\pm$0.03       & 0.44$\pm$0.02       & 4.64$\pm$0.06  & 1.32$\pm$0.03      & 0.80$\pm$0.02      & 0.56$\pm$0.01  &  166.4    \\
 \ion{H}{II}-7 & 7.55$\pm$0.21  & 8.10$\pm$0.20       & 2.37$\pm$0.08      & 27.34$\pm$0.61 & 8.00$\pm$0.15      & 4.26$\pm$0.08      & 3.13$\pm$0.08  &  139.3    \\
\hline
\end{tabular}
\caption{Energy fluxes (in units of $10^{-15}$ erg/cm$^2$/s) measured for the main emission lines identified in each \ion{H}{II} region spectra.}
 \label{tab:no2}
\end{table*}

\begin{figure*}[h!]
    \centering
    \includegraphics[width=\columnwidth]{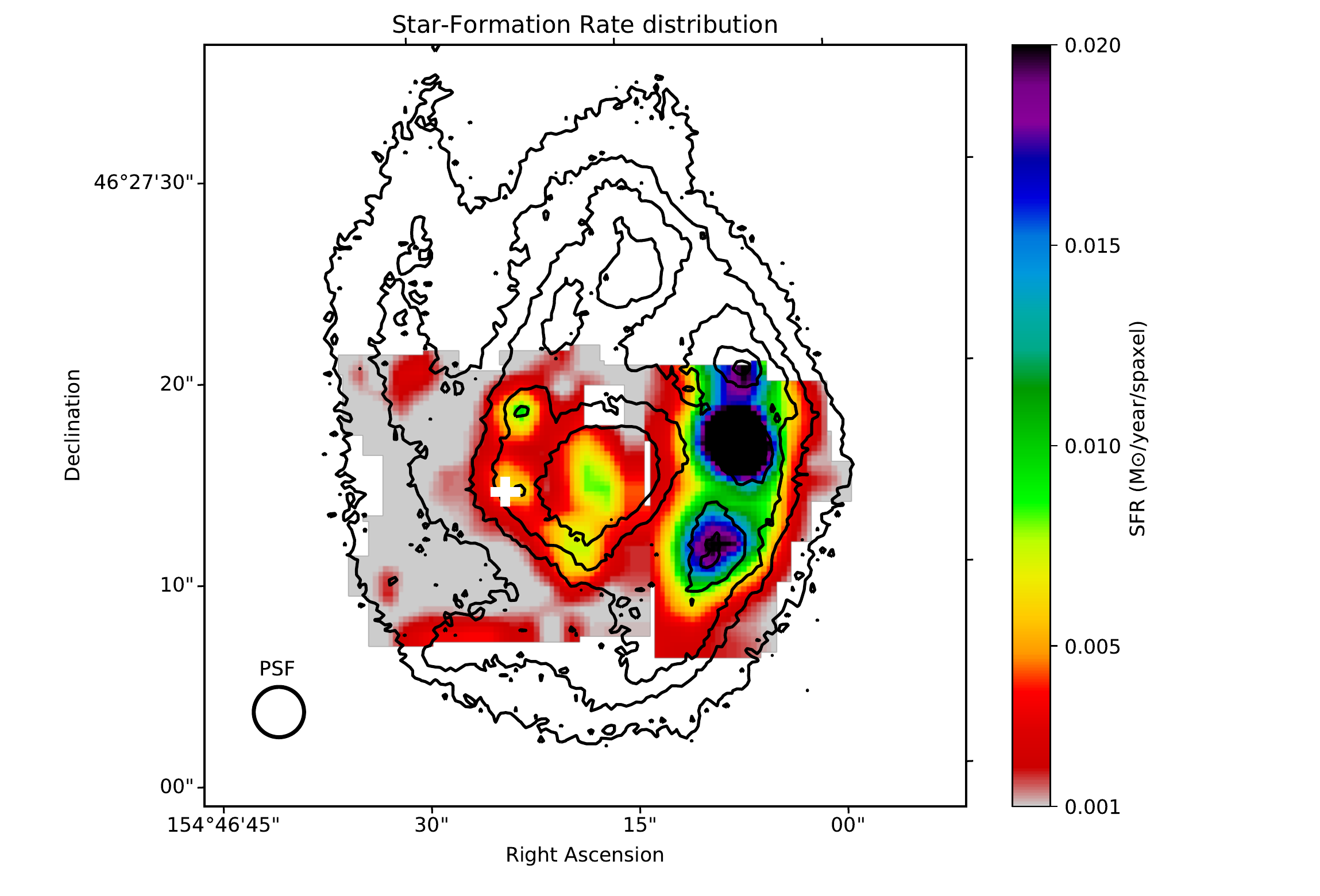}
    \includegraphics[width=\columnwidth]{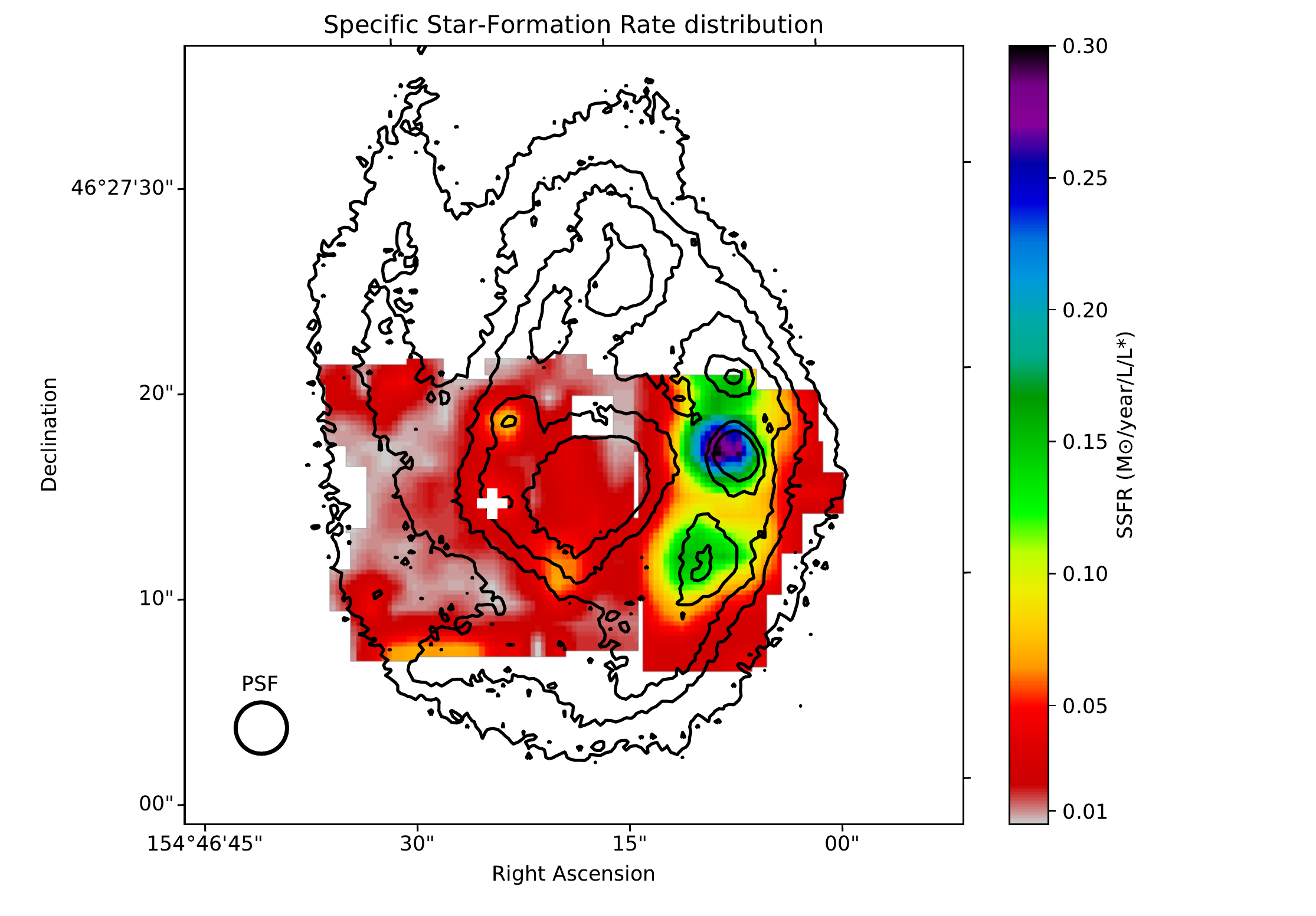}
    \caption{\textit{Left panel}: SFR map in the galaxy for the two fields covered by PMAS observations, computed using the \cite{Kennicutt1989} diagnostic.  \textit{Right panel}: The specific SFR obtained by weighting the SFR map and the rest-frame $B$-luminosity map, applying the same method described in \cite{Christensen2004}. The dimensions of each spaxel have been interpolated in the figure, showing elements of $0.25\times0.25$ arcsec. The SLSN position is marked with a white cross.}
    \label{fig:no2}
\end{figure*}

After correcting the data sets we finally derive several spatially-resolved physical properties of the host galaxy. We determine the star-formation rate (SFR) using the H$\alpha$ line flux F$_{H\alpha}$ diagnostic \citep{Kennicutt1989}, where: SFR[M$_{\odot}$yr$^{-1}]=7.9\times10^{-42}$ 4 $\pi$ d$_l^2$ F$_{H\alpha}$ (d$_l$ is the luminosity distance of the galaxy). The SFR in each spaxel is shown in Fig. \ref{fig:no2}. We find a low value for the SFR of the \ion{H}{II} region at the SLSN location of SFR$_4=0.22\pm0.01$ M$_{\odot}$yr$^{-1}$, while the highest values are found for the bright and younger regions 5 and 7: SFR$_5=1.43\pm0.02$ M$_{\odot}$yr$^{-1}$ and SFR$_7=0.78\pm0.02$ M$_{\odot}$yr$^{-1}$ (see also Table \ref{tab:no3}). The integrated SFR for the part of the galaxy covered by our observations is $\sim4.0\pm0.07$ M$_{\odot}$yr$^{-1}$. Assuming that the remaining part of the galaxy, that is smaller in area than the one covered by our observations, shows a similar value we estimate an integrated SFR for the total galaxy of $\sim7$ M$_{\odot}$yr$^{-1}$, in line with the SFR value inferred from the Galaxy Evolution Explorer (GALEX) near-UV data by \citet{Nicholl2017}. The authors of this paper also claim that $\sim70$\% of the total SFR in the galaxy is obscured and obtain an estimate of the total SFR using additional archival measurements in the InfraRed, X-ray and radio of SFR$_{tot}\simeq15$ M$_{\odot}$yr$^{-1}$. Our estimate is in agreement with their results.

Furthermore, we compute the specific SFR (sSFR), i.e., the star-formation rate per unit luminosity (sSFR$=$SFR/(L/L$^* $)). We use the method described in \citet{Christensen2004}, where a map of the rest-frame $B$-luminosity is weighted with the SFR map obtained above. The $B$-band rest-frame magnitude is derived from archival $g^\prime$-band Pan-STARRS images, which are almost identical to the rest-frame $B$-band wavelength range, and from this estimate the luminosity is obtained by computing the corresponding absolute magnitude after projecting the PMAS data onto the Pan-STARRS image. Finally, we multiply the SFR map with the ratio between the absolute $B$-magnitude map and the magnitude of a M$_B=-20.1$ galaxy (an average value that has been inferred for the break in the Schechter luminosity function at redshift $z=0.04$ and $z=0.07$ from the analysis of blue galaxies in the DEEP2 and COMBO-17 surveys, \citealt{Faber2007}). The results are shown in Fig. \ref{fig:no2}.

\begin{figure*}[ht!]
    \centering
    \includegraphics[width=\hsize]{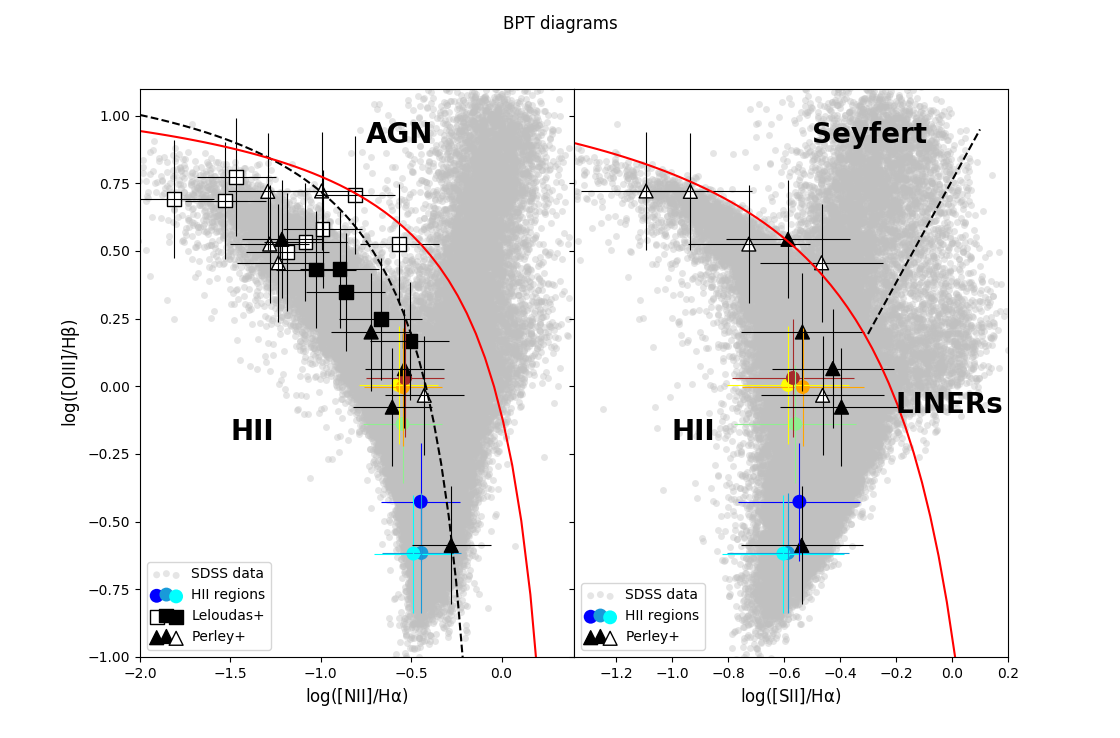}
    \caption{The BPT emission-line diagnostic used to distinguish the ionization mechanism acting in the nebular gas in galaxies. Two distinct diagnostics are used here: the left side shows [\ion{O}{III}] 5008 \AA\, / H$\beta$ versus  [\ion{N}{II}] 6584 \AA\, / H$\alpha$, while the right sides shows the [\ion{S}{II}] 6718/32 \AA\, doublet instead of [\ion{N}{II}]. The solid red curves in both panels denote the theoretical demarcation line between AGNs and star-forming galaxies proposed by \citet{Kewley2001}. In the left panel, the black dashed curve corresponds to the empirical demarcation between AGNs and SF-galaxies proposed by \citet{Kauffman2003}, while in the right panel the  black dashed line marks the division between active Seyfert galaxies and LINERs proposed by \citep{Kewley2006}. Grey data points are taken from the SDSS DR7 sample of galaxies described in \citet{Tremonti2004}. Coloured circles correspond to the \ion{H}{II} regions described in the text with the same colour coding used in Fig. \ref{fig:no1}. Squares and triangles represent SLSN-host data from \citet{Leloudas2015} and \citet{Perley2016}, empty symbols represent H-poor SLSNe, filled ones H-rich SLSNe. The SLSN 2017egm location (light-green circle) shows it is one of the most metal-rich Type-I SLSN.}
    \label{fig:no3}
\end{figure*}

The peculiarity of the host of SN 2017egm is underlined by its placement in the BPT diagram \citep{Baldwin1981}, which is a diagnostic based on emission-line flux ratios, and is used to distinguish star-forming galaxies from active galactic nuclei (AGN) like Seyferts and LINERs. The location on the star-forming branch is also an indicator of the ionization field of the gas, with large ionized \ion{H}{II} regions found in the upper left region of the BPT diagram \citep{Sanchez2015}. We use the emission-line fluxes measured in the integrated \ion{H}{II} regions to determine their respective location in the BPT diagrams. For comparison, we include two large samples of SLSN host galaxies from \citet{Leloudas2015} and \citet{Perley2016} distinguishing H-poor (Type I) and H-rich (Type II) SLSNe.

As shown in Fig. \ref{fig:no3}, all \ion{H}{II} regions of NGC 3191 analysed in this paper are located in the lower part of the star-forming branch of the BPT diagram, where metal-rich and low-ionization \ion{H}{II} regions and the bulk of the general SDSS galaxy population are found. The location of SLSN 2017egm is one of the lowest on the star-forming side of the BPT diagram, implying that SLSN 2017egm is one of the metal-richest H-poor SLSN found to date, more similar to the environment of H-rich SLSNe.

\begin{figure}[ht!]
    \centering
    \includegraphics[width=8.6cm]{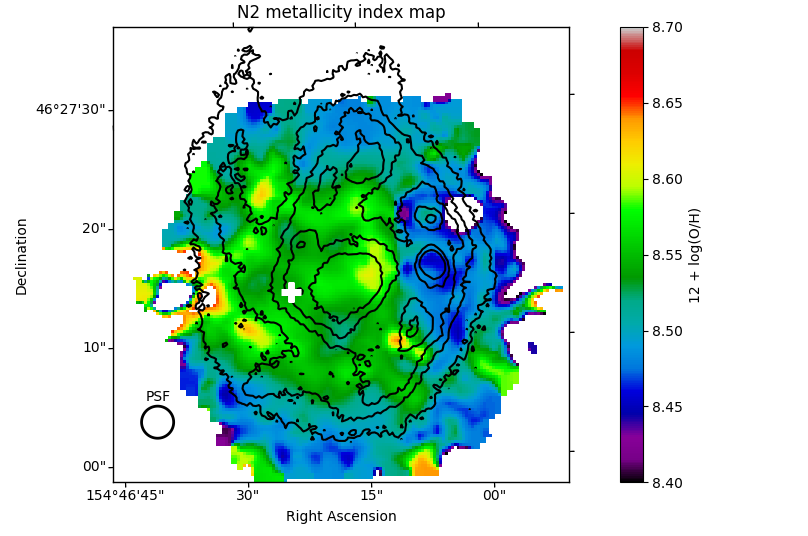}
    \caption{Metallicity map obtained from MaNGA data using the N2 indicator as given by \citet{Marino2013}. The location of SLSN 2017egm is reported with a black cross while the location of PTF10bgl is reported with a black circle.}
    \label{fig:no2b}
\end{figure}

For the estimate of other physical properties we used \texttt{PYNEB} which computes the physical conditions of \ion{H}{II} nebulae from specific diagnostic line ratios.
For the estimate of the metallicity we applied two methods. First, we used two emission-line indicators with the calibrations provided in \citet{Marino2013}: 1) the N2 method defined as 12+log(O/H) $=8.743+0.462\,\times\,$log([\ion{N}{II}]6584/H$\alpha$), and 2) the O3N2 method defined as 12+log(O/H)$= 8.533-0.214\,\times\,$log([\ion{O}{III}] 5007/H$\beta\,\times\,$H$\alpha$/[\ion{N}{II}]6584). The results for each \ion{H}{II} region are reported in Table \ref{tab:no3}. Thanks to the presence of auroral lines in some of the spectra, we were also able to obtain elemental abundances obtained by means of the ``direct'' method (see Table \ref{tab:App3}). Both methods indicate a higher metallicity for the centre of the galaxy ($Z_2=0.7Z_{\odot}$, with $Z_{\odot}=8.69$ \citep{Asplund2009}), while the SLSN site shows slightly lower metallicities ($Z_4=0.6Z_{\odot}$) similar to the bright \ion{H}{II} regions in the western part of the galaxy. In general, the metallicity distribution in the galaxy shows a decreasing distribution toward the external bright \ion{H}{II} regions, see also Fig. \ref{fig:no2b}, but has some asymmetry between the western and the eastern spiral arms.

Note that \cite{Nicholl2017} and \cite{Bose2017} find a much higher value for the metallicity using the BOSS spectrum of the galaxy centre of 2.0 and 1.6 Z$_\odot$ respectively. All these estimates of them use distinct R23-based metallicity indicators for estimating the metallicity, which however is known to give overestimates of Z \citep{Kennicutt2003,Yin2007}. In fact, from the same spectra we obtain a metallicity of 0.7~Z$_\odot$ using the N2 parameter from \cite{Marino2013}. Our estimate not only is based on the most updated calibrations available, but it is also supported by the spectral energy distribution modelling made in \citet{Nicholl2017}. \cite{Chen2017c} find a high metallicity value ($1.3\textnormal{Z}_\odot$) using the older calibration \citep{PettiniPagel2004} of the O3N2 indicator. However, they use the same analysis developed by \cite{Bianco2016}, which consists in checking a set of different metallicity indicators, and indeed find lower  values when considering the most updated calibrations, in agreement with our results.

We also use the [\ion{S}{II}] doublet to estimate the ionization using the formulation given in \citet{Diaz2000}, obtaining that the \ion{H}{II}-5,7 regions show a higher degree of ionization with respect to the other regions, including the one at the centre of the galaxy (see Table \ref{tab:no3}). [\ion{S}{II}] lines, in addition to other methods, have been also used for estimating the electron density and temperatures of the gas in \ion{H}{II} regions: in general we find electron density values ($\sim100$ cm$^{-3}$) and temperatures ($T_e \sim 10000$ K) typical of \ion{H}{II} regions, although for the central regions we observe lower values ($T_{e,2} \sim 7500$~K).

\begin{table*}
    \centering
    \begin{tabular}{lcccccc}
\hline
 Region                  & SFR  & sSFR/L/L$^*$ & N2   & O3N2         & log U  & E(B-V)  \\
\hline
                   & $(M_{\odot}yr^{-1})$  & $(M_{\odot}yr^{-1})$  & 12 + log(O/H)   & 12+ log(O/H)    &     
    &  (mag)   \\
\hline
 \ion{H}{II}-1 &  0.26$\pm$0.01 & 0.56$\pm$0.15 & 8.54$\pm$0.16 & 8.53$\pm$0.18    & -3.35$\pm$0.01 &  0.37$\pm$0.03\\
 \ion{H}{II}-2  & 0.56$\pm$0.02 & 0.73$\pm$0.15 & 8.54$\pm$0.16 & 8.57$\pm$0.18    & -3.58$\pm$0.01 &  0.46$\pm$0.03\\
 \ion{H}{II}-3 &  0.41$\pm$0.01 & 0.41$\pm$0.17 & 8.52$\pm$0.16 & 8.56$\pm$0.18     & -3.46$\pm$0.01 &  0.41$\pm$0.01\\
 \ion{H}{II}-4 &  0.22$\pm$0.01 & 0.42$\pm$0.14 & 8.49$\pm$0.16 & 8.45$\pm$0.18    &  -3.35$\pm$0.01 &  0.30$\pm$0.03\\
 \ion{H}{II}-5 &  1.43$\pm$0.02 & 2.67$\pm$0.36 & 8.48$\pm$0.16 & 8.41$\pm$0.18    &  -3.01$\pm$0.01 &  0.22$\pm$0.03\\
 \ion{H}{II}-6 &  0.25$\pm$0.03 & 0.96$\pm$0.23 & 8.49$\pm$0.16 & 8.42$\pm$0.18   &  -3.10$\pm$0.01 &  0.16$\pm$0.03\\
 \ion{H}{II}-7 &  0.78$\pm$0.02 & 1.93$\pm$0.30 & 8.50$\pm$0.16 & 8.41$\pm$0.18   & -2.99$\pm$0.01 &  0.24$\pm$0.03 \\
\hline
\end{tabular}
\caption{Physical properties derived from the integrated spectra of each \ion{H}{II} region: (2nd column) the star-formation rate and (3rd) the specific star-formation rate per unit luminosity, (4th-5th) the metallicity obtained with the N2 and the O3N2 indicators, (6th) the ionisation parameter log $U$ estimated using the [\ion{S}{II}] lines and the formulation given in \cite{Diaz2000}, and (7th) the estimate of the extinction $E(B-V)$. }
\label{tab:no3}
\end{table*}

\section{Stellar populations and kinematics shaped by an interaction with its neighbour?}

\begin{figure}[h!]
    \centering
    \includegraphics[width=8.6cm]{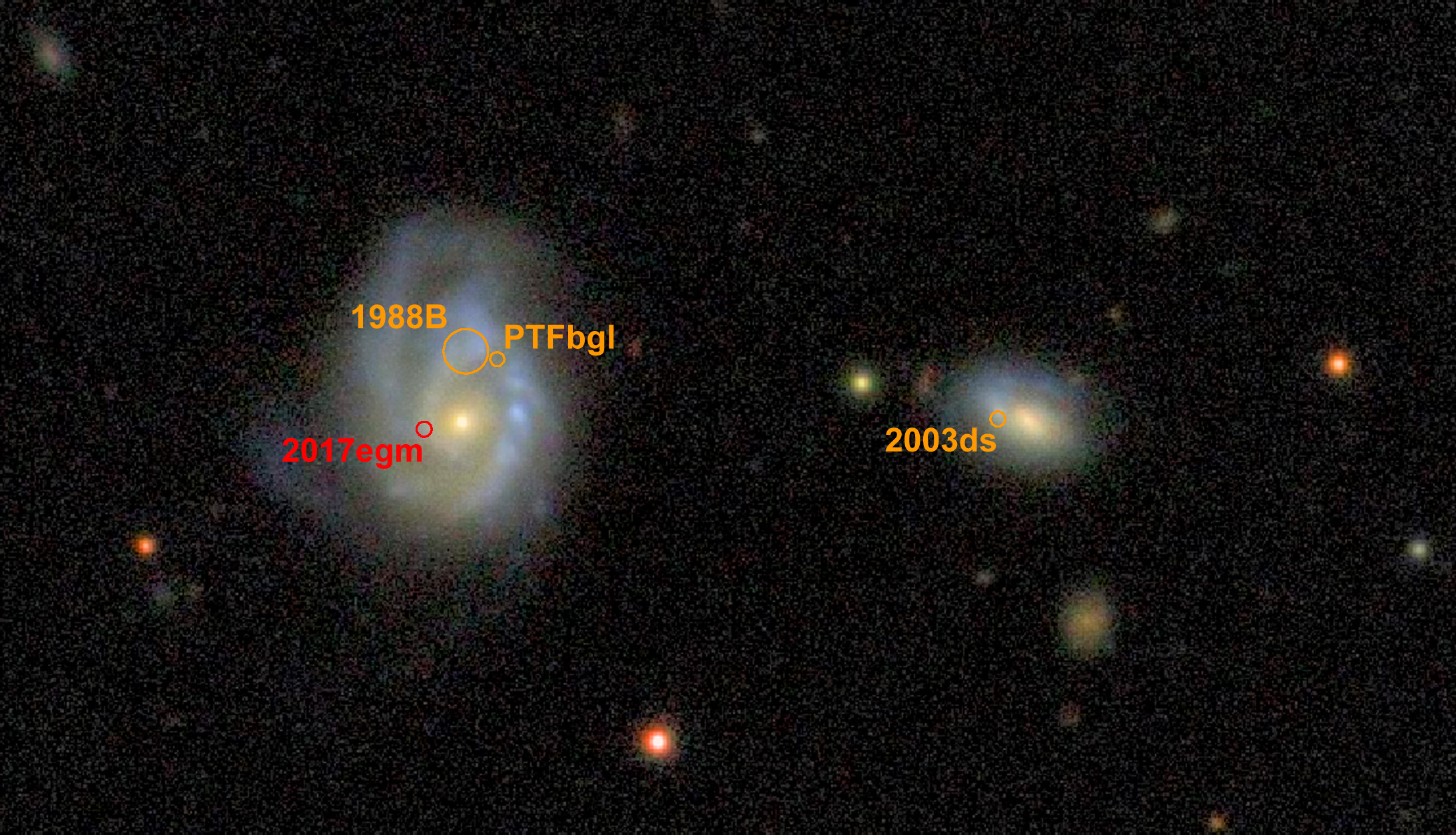}
    \caption{Colour image of NGC 3191 and its neighbour MCG+08-19-017 created combining multiband data from the PanSTARRS and the SDSS image catalogues. We also mark the locations of all four SNe that occured in this galaxy pair. North is up and East is to the left, the field of view is $3\farcm2\times1\farcm8$.}
    \label{fig:no10}
\end{figure}

\begin{figure}
    \centering
    \includegraphics[width=8.6cm]{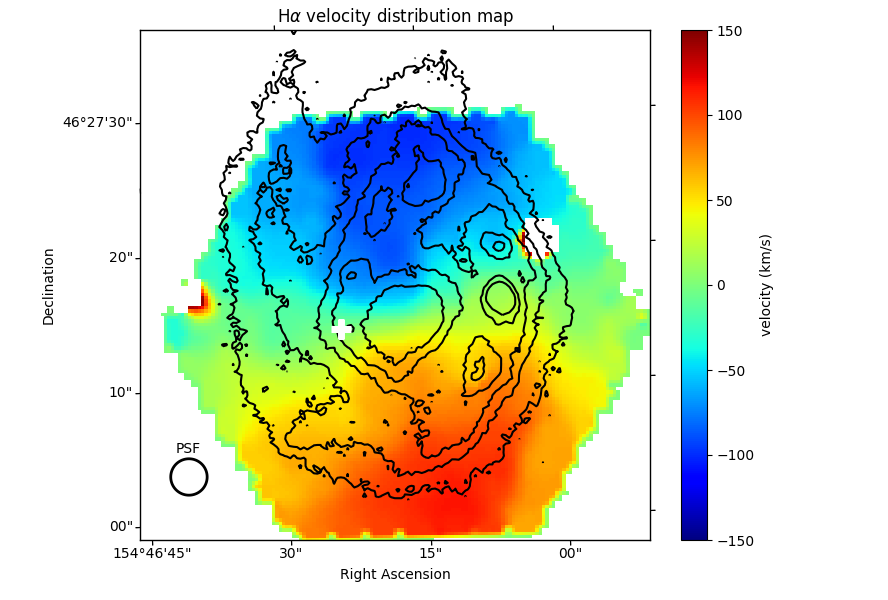}
        \includegraphics[width=8.6cm]{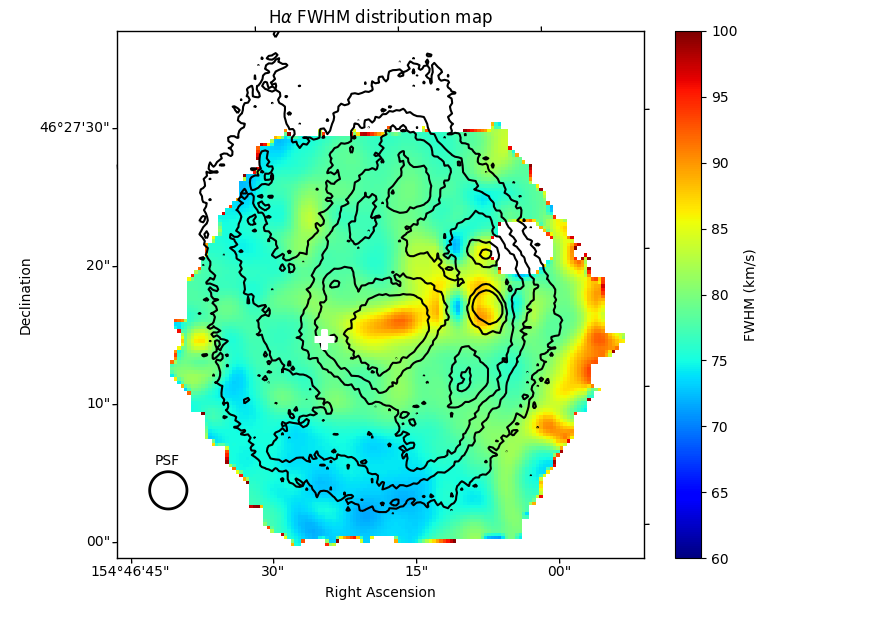}
    \caption{\textit{Upper panel}: Velocity field map of the gas in NGC 3191 with kinemetry ellipses overplotted as dashed lines. \textit{Lower panel}: FWHM H$\alpha$ distribution map.}
    \label{fig:no11}
\end{figure}

\begin{figure}[ht!]
    \centering
    \includegraphics[width=9.0cm]{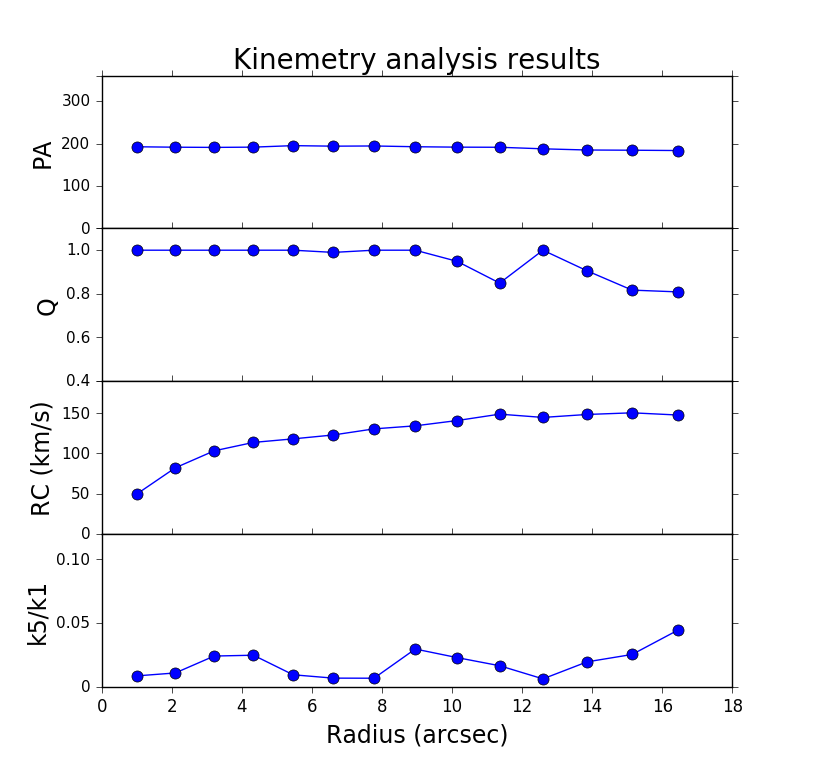}
    \caption{Best-fit parameters obtained from the kinemetry analysis \citep{Krajnovic2006} for different positions of the tilted-rings relative to the centre of NGC 3191. From the top: Position angle in degrees, the ellipticity parameter q,  the rotation curve (RC) that corresponds to the first order $k1$ of the harmonic expansion corrected for the inclination of the galaxy, and the $k5/k1$ ratio.}
    \label{fig:no11a}
\end{figure}

NGC 3191 is not an isolated galaxy. It has a companion, MCG+08-19-017, at a projected distance of $\sim45$ kpc (between the galaxy centres) and a radial velocity difference of only $\sim200$ km/s. A colour picture of the two galaxies is show in Fig. \ref{fig:no10}. From the analysis of the SDSS spectra described in Section 2, we find that MCG+08-19-017 has a low SFR of 0.15 M$_\odot$ yr$^{-1}$ and an older stellar population as shown by a low H$\alpha$ EW of only 21 \AA{} and a significant stellar absorption component in the H$\beta$ line (EW of 6.3 \AA{}). The metallicity is similar to the values observed in the centre of NGC 3191 with $Z(N2) \sim0.7$  M$_{\odot}$ using the calibrations of \cite{Marino2013}\footnote{The SDSS spectrum of the galaxy companion can be found at https://dr12.sdss.org/spectrumDetail?mjd$=$ 52614\&fiber$=$505\&plateid$=$944}. No IFU data are available for the companion. 

\begin{figure}[ht!]
    \centering
    \includegraphics[width=8.6cm]{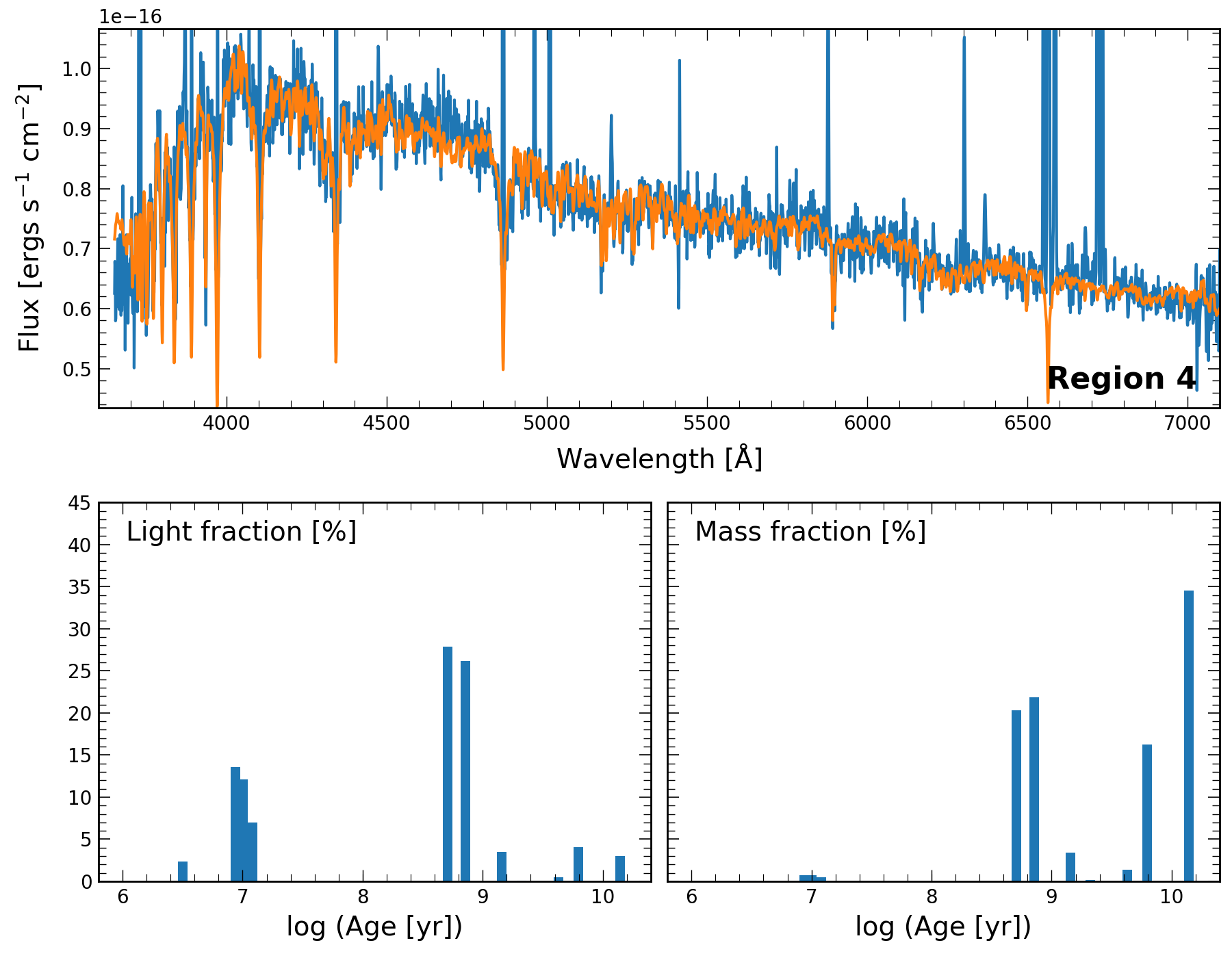}
    \includegraphics[width=8.6cm]{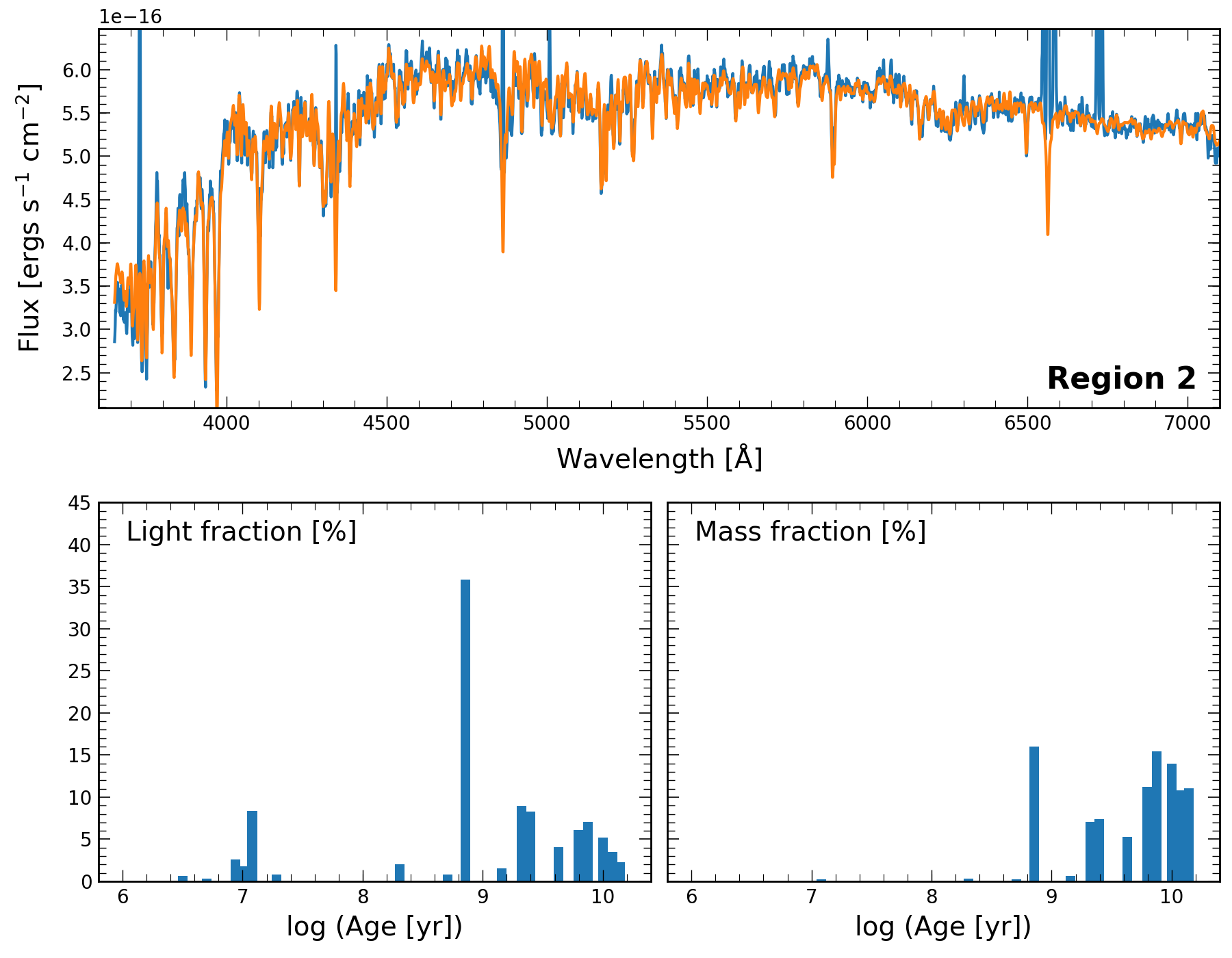}
    \includegraphics[width=8.6cm]{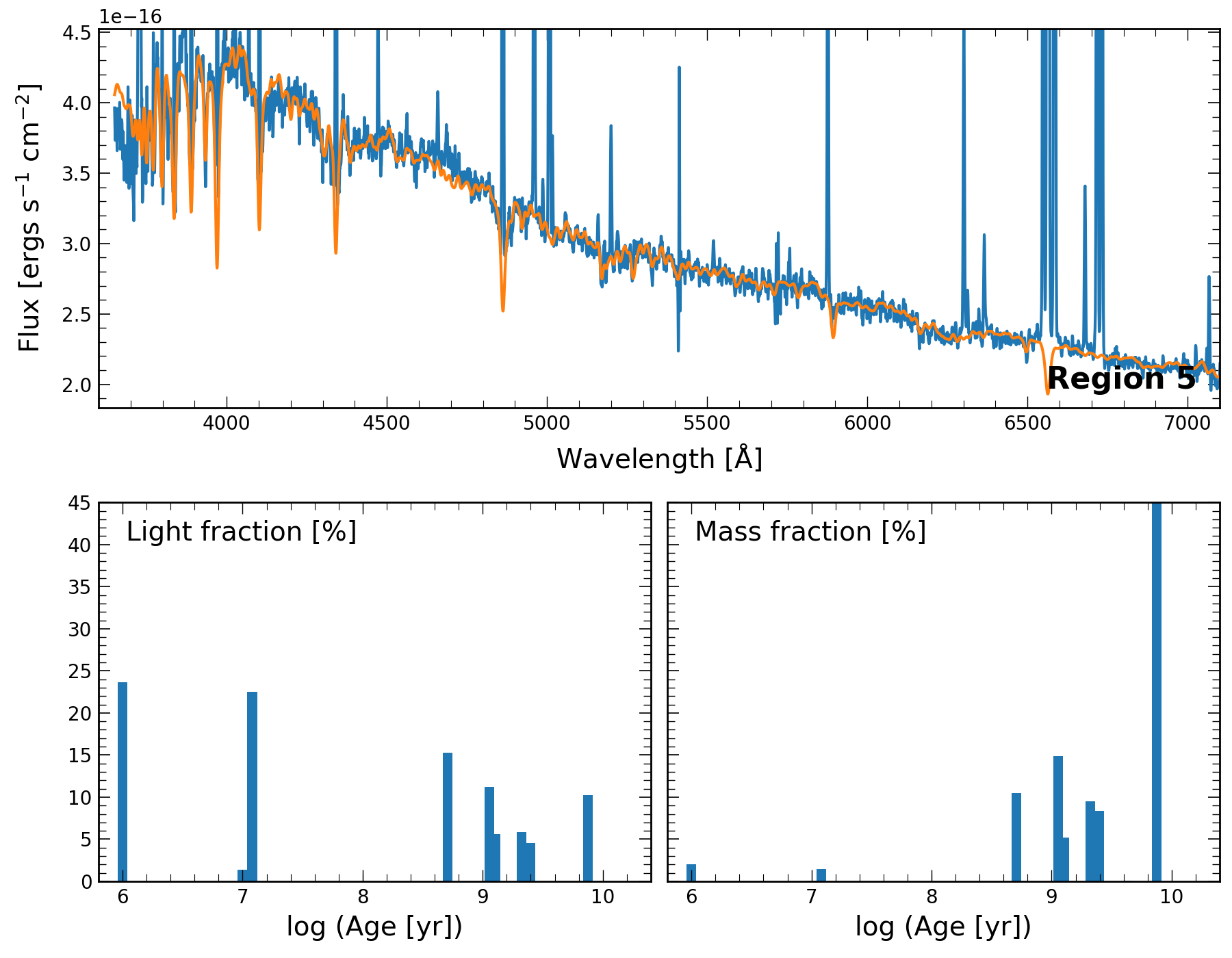}
    \caption{From top to bottom, the stellar population analysis of the \ion{H}{II} regions number 4 (the SLSN region), number 2 (the galaxy centre) and number 5 (the most active region in the galaxy), respectively. Upper panels show the fit to the observed spectral energy distributions, while lower panels show the fraction in luminosity of stellar populations with different ages (left) and their contribution to the total stellar mass (right).}
    \label{fig:no11c}
\end{figure}

The somewhat disturbed morphology of NGC 3191 suggests that the two galaxies could have had interacted gravitationaly, which would explain the observed distribution of SFR and metallicity in NGC 3191. In fact, MCG+08-19-017 also shows more intense star-formation on its Eastern side, in the direction towards NGC 3191, hence suggesting that some near pass-by of the two galaxies could have compressed gas in both on their respective sides of the closest encounter. In order to investigate the kinematics in NGC 3191, we use the observations available from the MaNGA survey. In Fig. \ref{fig:no11} we plot the velocity field determined from fitting the H$\alpha$ line with a Gaussian and taking the peak and FWHM for each spaxel. The velocity field of the host is surprisingly regular considering its morphology, only the outer parts of the disk seem to be somewhat twisted.  The width of the lines only show an enhancement in the central region and within the brightest of the three large SF regions in the West. Line broadening in the center is expected as it probes a higher amount of gas in the bulge. The enhancement in a active SF region is a reflection of turbulence due to the ongoing heavy SF.

To quantify disturbances in the velocity field we perform a kinemetry analysis following the method described in \cite{Krajnovic2006} and plot the results in Fig. \ref{fig:no11a}. From the ratio of k5/k1 we see that there is no strong disturbance in a smooth velocity field which would lead to much higher values for k5 (the first higher-order term in the analysis). However, the ellipticity parameter (q) increases  from near-circles (as expected for a near-face-on galaxy) in the centre to more elongated ellipses in the outer parts of the host, while in undisturbed galaxies the ellipticity is expected to be rather constant. This could be a sign of interaction with its neighbour, introducing a tangential or warp-like disturbance in the velocity field.

Considering the evidence for interaction and the rather atypical location of the SLSN next to the centre, we also perform a thorough analysis of the stellar population in the galaxy. For this we use the \texttt{STARLIGHT} code \citep{CidFernandes2005} which estimates the properties of the stellar population from the observed spectral energy distribution, and using the stellar-population synthesis models as given in \citet{Bruzual2003}. We use the spectra of the \ion{H}{II} regions extracted from the MaNGA datacube which have better spectral resolution and a wider wavelength range. The total stellar masses and distributions of the stellar mass with age for each \ion{H}{II} region are listed in Table \ref{tab:no4}. In Fig. \ref{fig:no11c} we show the detailed fitting and stellar population distribution for three particular regions: 1) the SLSN region, 2) the region (\ion{H}{II}-2) located at the galaxy centre and 3) the \ion{H}{II}-5 region, the currently most active SF region in the galaxy.

\begin{figure}[ht!]
    \centering
    \includegraphics[width=8.6cm]{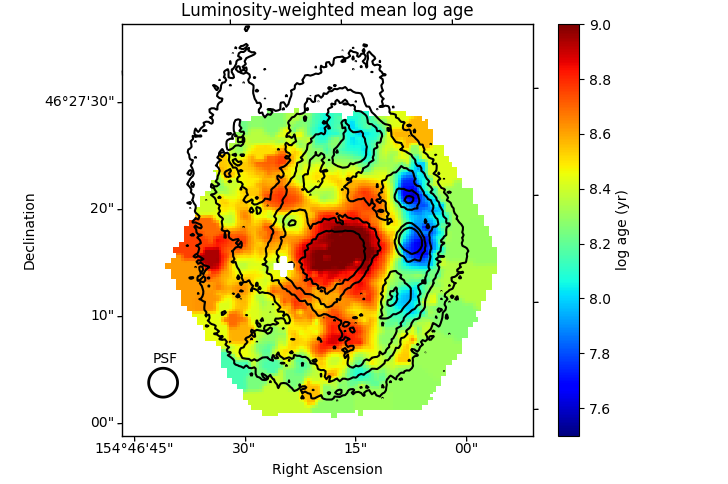}
    \caption{The luminosity weighted (log) age distribution estimated with the \texttt{STARLIGHT} code, following the prescriptions given in \citet{CidFernandes2013}. The location of SLSN 2017egm is reported with a black cross while the location of PTF10bgl is reported with a black circle.}
    \label{fig:no12}
\end{figure}

We note that \ion{H}{II} regions characterised by the youngest ages (weighted by their luminosities) show the lowest metallicities. This becomes clear when looking at the distribution of the age weighted by the luminosity in the galaxy\footnote{We run \texttt{STARLIGHT} on the MaNGA cube using the same SSPs ingredients as the \ion{H}{II} spectra. The output was packed and processed with the {\sc p}y{\sc casso} pipeline \citep{CidFernandes2013,Amorim2017}.} in Fig. \ref{fig:no12}, with the distribution of the metallicity estimated with the N2 indicator (see Fig. \ref{fig:no2b}). Our detailed stellar-population analysis shows the presence of at least two distinct stellar populations in all \ion{H}{II} regions, one spanning from $1-10$ Gyrs, which contributes almost all of the stellar mass, and an additional, very young population ($1-10$ Myrs), see also Table \ref{tab:no4}. While the high luminosity of the most active SF regions on the western side of the galaxy is due to an additional very young population of around 1 Myr (see also the lower panel in Fig. \ref{fig:no12}), the galaxy centre is dominated by the old ($>1$ Gyr) stellar population with only very low SF activity in the past tens of Myrs (middle panel in Fig. \ref{fig:no12}). The observed light from SF regions close to the centre including the SLSN site have an older population at the lower end of the initial SF activity at $\sim$ 1 Gyr down to several 100~Myrs ago, but additional new activity started at 10~Myrs and extends down to $2-4$ Myrs. A stellar population of 3--10~Myr would correspond to the evolution time of a massive stars of $20-60$ M$_\odot$ until explosion \citep[see e.g.][]{Kunca13a}. The fraction in luminosity of stars younger than 30~Myr is high for \ion{H}{II} regions 4-7, but these regions are also characterised by a low metallicity value, see Table \ref{tab:no3}. Consequently, we infer that the presence of these young stars is responsible for the observed lower metallicity in these regions, and that these young and massive star populations are characterised by an intrinsically lower metal abundance ($<0.6Z_{\odot}$) than the entire host, including the progenitor of SLSN 2017egm.

All \ion{H}{II} regions show a SF peak around 10--15 Myrs. At that time, a burst of star-formation very likely involved significant galaxy. These metal-poor young populations could very well come from an interaction with the neighbouring galaxy, leading to compression and shocks of the gas in the part of the host facing the companion. Possibly, the same process lead to enhanced star-formation on the Eastern part of the neighbour galaxy. The slightly ``older'' regions of this starburst around the center of NGC 3191 maintain the information on the burst of star-formation caused by this gravitational interaction. The SLSN progenitor could be the first product of this prolonged (few Myr) burst of star-formation while the recently triggered SF in the western \ion{H}{II} regions are still too young to actually host any SNe. In this picture, the star-formation would still be taking place with older metal-enriched gas, hence still posing a problem if the SLSN has to be in a metal-poor environment. Such interactions, however, can also funnel metal-poor \ion{H}{I} gas into the central regions of a galaxy, thereby giving rise to a SLSN in a metal-rich environment. Another possibility is transfer of metal-poor gas from the interacting galaxy to NGC 3191. However, the spectrum of the centre of MCG+08-19-017 shows a metallicity value similar to the one of NGC 3191, making this possibility less likely. Instead of a direct gas transfer, pristine gas from possible residuals of former dwarf companions ripped apart by the interaction, could also be transferred into the main galaxy. Resolved HI data of both galaxies could give some clues on this scenario \citep{Michalowski2015}. 

\begin{table*}
    \centering
    \begin{tabular}{lcccccc}
\hline
 Region        & M$_*$             & Age & $f(age<30 Myr)$   & $f(30Myr<age<Gyr)$          & $f(age>Gyr)$  & $f(L<30Myr)$  \\
\hline
         &    $(log M_{\odot})$        &  $(log [yr])$   & $(\%)$   & $(\%)$          & $(\%)$   &  $(\%)$   \\
\hline
 \ion{H}{II}-1 & 8.76 & 8.43  & 1.4 & 37.9 & 60.7 & 23.9 \\
 \ion{H}{II}-2 & 9.54 & 8.43  & 0.4 & 16.7 & 82.9 & 14.5  \\
 \ion{H}{II}-3 & 9.12 & 8.69  & 0.9 & 24.3 & 74.8 & 20.7 \\
 \ion{H}{II}-4 & 8.28 & 8.24  & 2.1 & 42.2 & 55.7 & 35.0 \\
 \ion{H}{II}-5 & 8.86 & 7.93  & 3.5 & 10.5 & 86.0 & 47.4 \\
 \ion{H}{II}-6 & 7.83 & 7.81  & 4.3 & 30.1 & 65.6 & 47.1  \\
 \ion{H}{II}-7 & 8.77 & 8.07  & 3.1 & 36.5 & 60.4 & 38.5 \\
\hline
\end{tabular}
\caption{Physical properties for each \ion{H}{II} region from \texttt{STARLIGHT}: total stellar mass (2nd column), the light-weighted age  (3rd), the mass fraction of stellar populations younger than 30 Myr, between 30 Myr and 1 Gyr and older than 1 Gyr, respectively (4th-6th) and finally the fraction in luminosity of stellar populations younger than 30 Myr (7th). }
\label{tab:no4}
\end{table*}

\section{The locations of SN 1988B, SN 2003ds and PTF10bgl}
The galaxy pair NGC 3191 and MCG+08-19-017 have hosted not only SN 2017egm but another three SNe in the past two decades. This  underlines the increased SF and SN activity due to mutual interaction in both galaxies as concluded in the previous section. In Fig. \ref{fig:no10} we plot the locations of all four SNe in the galaxy pair.

SN 1988B was a Type Ia SN discovered by P. Wild \citep{IAUC4533} and classified as a SN Ia \citep{IAUC4535}. Unfortunately, there are no precise enough coordinates reported in the literature, but \citep{IAUC4533} give the location as 10\farcs0 north of the galaxy center. \citep{IAUC4549} also mention the presence of H$\alpha$ emission on top of the SN spectrum, indicative of a SF region at the SN site. SNe Ia can occur in any kind of galaxy and requires an older population (generally from $2 \times 10^8$ to $10^{10}$ years, \citep{Maoz2012} ) for the white-dwarf progenitor to form. NGC 3191 does have a large underlying older population with ages of 1--10Gyr, hence the occurrence of a SN Ia in this galaxy is not surprising.

SN 2003ds was of Type Ic and took place in the neighbour galaxy. Its location near the bright SF region to the West which we attribute to the mutual interaction between the two galaxies matches well with expectations: Type Ic SNe are presumably Wolf-Rayet stars that lost their H and He envelope through winds or smaller outbursts \citep[e.g.][]{Georgy2009, Tramper2015}, hence the original progenitor had to have been a rather massive star of $>$25--30 M$_\odot$. An alternative is a binary progenitor where the companion has removed the envelope of the star, leaving a relatively small progenitor star. By studying the age of surrounding HII regions, \citep{Kunca13a} found a preference for young ages and therefore the single-star scenario, albeit with some exceptions. No reliable Type Ic progenitor detection has been made to date. Ic SNe are presumably requiring around solar metallicity  which facilitate the strength of stellar winds \citep[see e.g.][]{Leloudas11, Modjaz11, Kunca13a} but the statistical significance of the results are controversial \citep{Leloudas11}. Moreover, all agree that the metallicities are higher than those found for GRB (and hence SLSN-I) environments. MCG+08-19-017 has, indeed, a similar metallicity as the core of NGC 3191, however the Legacy Survey spectra only probe the central part of the galaxy and not the SF region of 2003ds. Since there are no IFU data on that galaxy, we cannot draw any further conclusions on the properties of the progenitor site at this point.

PTF10bgl is the most surprising of the three previous SNe as it was classified as a SLSN Type II \citep{Arcavi2010, Bose2017}, hence making NGC 3191 both the host of a SLSN Type I and Type II, which is so far unprecedented. PTF10bgl was located just north of the three major SF regions in the West in the same spiral arm but in a ``gap'' between SF regions. Fortunately, this region is also covered by the MaNGA data. The metallicity at the site of 10bgl is surprisingly similar to the site of 2017egm with 12+log(O/H)$=$8.52 $\pm$0.18 (N2 parameter) and a log age of 8.32 yr. The location of PTF10bgl, however, is outside any active \ion{H}{II} region in the host galaxy, which reflects in a much lower SFR measured at the SLSN site, SFR=$0.05\pm0.01$ M$_{\odot}$yr$^{-1}$. These results show that the environments of Type-I and Type-II SLSNe could have very similar physical properties.

\section{Conclusions}

In this work we present IFU data of a large part of the SLSN-2017egm host galaxy, NGC 3191, which is the closest SLSN-I discovered to date and has hosted another two SNe in the past, among them a SLSN-Type II. NGC 3191 is a large spiral galaxy with an integrated stellar mass of $M_*=10^{10.7}M_{\odot}$. The galaxy has an average metallicity of $\sim0.6$ Z$_\odot$, with a typical negative gradient distribution towards the external regions, as it is usually found in bright large spirals. However, these galaxies do not represent the canonical hosts for SLSNe Type I, which are usually low-mass star-bursting dwarf galaxies \citep{Leloudas2015, Schulze2016, Perley2016, Cikota2017}. The \ion{H}{II} region around the SLSN, with a size of $\sim 1$ kpc, has one of the highest metallicity values ($Z_4=0.6Z_{\odot}$) among the known SLSNe Type-I to date and higher than the proposed cut-off for SLSNe Type-I progenitors of $0.4-0.5Z_\odot$ \citep{Schulze2016, Chen2017b}. The only other IFU study of a SLSN-I host by \cite{Cikota2017} also show the SLSN not to be coincident with the lowest metallicity (and highest SFR) region, but the value at the site of PTF11hrq (12+log(O/H)$\sim$8.2) is still way below what we find for the site of SN 2017egm. Our metallicity value is based on the most recent calibration of the N2 and O3N2 indicators by \cite{Marino2013}. Taking  other, less reliable, but frequently used calibrations in the literature for comparison (see Table \ref{tab:App2}) we consistently find higher metallicities for the SLSN region. We also used the S23 indicator that uses the bright doublet lines of [\ion{S}{III}] $9069-9523\lambda\lambda$ in the parametrisation given by \cite{PerezMontero2005}. These lines are among the brightest in \ion{H}{II} regions, also at high metallicities, and do not suffer large extinction 
effects\footnote{See \cite{Garcia-Benito2010} for the first map of this parameter using IFU data.}. 

SLSN 2017egm is not closely tracing the SF in its host as is usually the case for SLSNe Type I \citep{Lunnan2015} and also for GRBs and their associated broad-line Type Ic SNe \citep{Fruchter2006,Blanchard2016}, however see the resolved study of PTF11hrq by \cite{Cikota2017}. The SFR value measured at the SLSN site from the H$\alpha$ flux ($SFR_4=0.22\pm0.01$ M$_{\odot}$yr$^{-1}$) is similar to values found for the integrated spectra of other SLSN hosts. Neither did SLSN 2017egm explode in the youngest region of its host. Our stellar population analysis shows the presence of two main stellar populations with ages of $>500$ Myr for the older population, and around 10 Myr for the younger population. In the SLSN region, the bulk ($54\%$) of the observed luminosity comes from an older population of stars, but a conspicuous fraction ($35\%$) originates from a young ($<30$ Myr) stellar population. In the bright SF regions in the western part of the galaxy a small fraction ($3\%$) of stars show an age less than 10 Myr. Additional evidences from other indicators, such as the observed EWs of Balmer lines and \ion{He}{I} 4471$\lambda$ \AA{} suggest the presence of stars with an age of the order of $\sim10$ Myr and possibly less. This value is larger than the value derived for the progenitor of PTF12dam \citep{Thoene2015}, which, however, was possibly youngest host to date, but similar to the age of PTF11hrq \citep{Cikota2017}. Still, this value is in line with the age of a rather massive star as required for current models of SLSNe Type I. 

Star-formation in this galaxy seems to have been at a low level for several 100 Myrs, after which it experienced a new recent burst in SF. Evidence for this is found in the brightest \ion{H}{II} regions. Their location, in the apparent direction toward the companion galaxy MCG+08-19-017, could suggest a gravitational interaction which triggered the recent star-formation. Kinematic analysis of NGC 3191 shows some deviations in the external regions of the galaxy with a ``twist'' in the velocity field that could very well be explained by an interaction within the last tens of Myrs. Evidence for an interaction can also be inferred from the metallicity distribution map, Fig. \ref{fig:no2b}, where a negative gradient toward the western side of the galaxy (where the youngest \ion{H}{II} regions are observed) is visible, but no such gradient exists on the eastern side. The same trend is visible in Fig. 2 of \citet{Chen2017c}, where the authors use distinct R23-based metallicity indicators, that as stated above overestimate the metallicity. Rearrangement of gas in the galaxy, gas transfer or the interaction causing metal-poor \ion{H}{I} gas to be funneled into the galaxy could explain the occurrence of a SLSN Type I at low metallicity even in this solar-metallicity galaxy and indeed there is a small population of metal-poor stars present at the SLSN site. New SF activity due to interaction was also proposed for three other SLSN-I hosts \citep{Cikota2017, Chen2017a}.

NGC 3191 has previously hosted two other SNe, a Type-Ia (SN 1988B) and a Type-II SLSN (PTF10bgl), while the neighboring galaxy MCG+08-19-017 hosted another Type-Ic SN (SN 2003ds). The presence of two SLSNe of different type in the same galaxy is a unique case so far. PTF10bgl shares very similar physical properties of the environment with 2017egm such as the gas metallicity and the stellar age. The low SFR value is likely due to its location, which is at the border of the most active \ion{H}{II} regions. We infer similar progenitor properties for both SLSNe from our analysis, although they belong to different SLSN types. We consider these evidences as important test-beds for SLSNe progenitor models.

Our study shows the importance of spatially resolved studies and detailed analysis of stellar populations to infer SN-progenitor properties. The detailed analysis of the spectra of different  \ion{H}{II} regions in the host allows us to understand the origin of the apparent metal-anomaly in the SLSN environment. The implications of this result are several: One has to be careful when inferring characteristics of SN progenitors from the physical properties of their unresolved host galaxies. While this provides valuable information in large statistical samples, it can be deceiving when used for individual cases. Moreover, a detailed analysis of the observed spectra is required, which takes into account the properties of the stellar population, not only those of the emitting gas. Finally, the use of powerful tools for studying galaxy kinematics are equally important to unveil traces of gravitational interactions that can trigger bursts of star-formation. A global analysis is then necessary to precisely constrain the physical characteristics of very massive stars on their way to explode as SLSNe.

\begin{acknowledgements}
We thank the referee for her/his constructive comments that have improved the paper. We also thank Steve Schulze and Yan Lin for their important comments to the paper. LI, CT, ZC, AdUP and DAK acknowledge support from the Spanish research project AYA2014-58381-P. CT and AdUP furthermore acknowledge support from Ram\'on y Cajal fellowships RyC-2012-09984 and RyC-2012-09975. DAK and ZC acknowledge support from Juan de la Cierva Incorporaci\'on fellowships IJCI-2015-26153 and IJCI-2014-21669. RGB acknowledges support from the Spanish Ministerio de Econom\'ia y Competitividad, through projects AYA2016-77846-P and AYA2014-57490-P. LI wishes to thank Anna Serena Esposito for her kind availability and support in organizing the figures presented in this paper.
\end{acknowledgements}

\bibliographystyle{aa}

\begin{appendix}
\section{Tables}

\begin{table*}
    \centering
    \begin{tabular}{lccccccc}
\hline
 Region        & [\ion{O}{II}] 3727,29           & [\ion{Ne}{III}] 3869   & [\ion{O}{III}] 4959   & \ion{He}{I} 5876          & [\ion{O}{I}] 6300   & [\ion{O}{II}] 7319,30   & [\ion{S}{III}] 9069,9531   \\
\hline
 \ion{H}{II}-1 & 2.40$\pm$0.07 & 0.18$\pm$0.04 & 0.25$\pm$0.06 & 0.21$\pm$0.04 & 0.19$\pm$0.05 & -- & 1.67$\pm$0.06  \\
 \ion{H}{II}-2 & 2.27$\pm$0.09 & 0.43$\pm$0.10 & -- & 0.27$\pm$0.11 & -- & -- & 2.10$\pm$0.20   \\
 \ion{H}{II}-3 & 2.40$\pm$0.68 & -- & -- & 0.26$\pm$0.06 & 0.26$\pm$0.08 & -- & 1.95$\pm$0.92  \\
 \ion{H}{II}-4 & 3.28$\pm$0.06 & 0.13$\pm$0.03 & 0.42$\pm$0.02 & 0.23$\pm$0.02 & 0.19$\pm$0.01 & 0.08$\pm$0.02 & 1.48$\pm$0.03  \\
 \ion{H}{II}-5 & 31.71$\pm$0.63 & 1.00$\pm$0.09 & 4.99$\pm$0.18 & 1.91$\pm$0.08 & 1.15$\pm$0.07 & 0.75$\pm$0.16 & 6.05$\pm$0.30   \\
 \ion{H}{II}-6 &  2.89$\pm$0.06 & 0.09$\pm$0.01 & 0.45$\pm$0.02 & 0.19$\pm$0.01 & 0.12$\pm$0.01 & 0.06$\pm$0.01 & 1.37$\pm$0.03   \\
 \ion{H}{II}-7 & 15.39$\pm$0.28 & 0.58$\pm$0.08 & 2.73$\pm$0.11 & 0.98$\pm$0.05 & 0.67$\pm$0.06 & 4.00$\pm$ 0.41 & 9.32$\pm$0.22    \\
\hline
\end{tabular}
\caption{Energy fluxes (in units of $10^{-15}$ erg/cm$^2$/s) measured for additional emission lines identified in each \ion{H}{II} region spectrum.}
 \label{tab:App1}
\end{table*}

\begin{table*}
    \centering
    \begin{tabular}{lcccccc}
\hline
 Region                  & N2 (M13)  & N2 (PP04) & O3N2 (M13)   & O3N2  (PP04)     & $R_{23}$ &  $S_{23}$ \\
\hline
 \ion{H}{II}-1 & 8.54$\pm$0.16 & 8.68$\pm$0.18 & 8.53$\pm$0.18 & 8.73$\pm$0.14 & 8.89$\pm$0.19 & 8.36$\pm$0.20 \\
 \ion{H}{II}-2 & 8.54$\pm$0.16 & 8.69$\pm$0.18 & 8.57$\pm$0.18 & 8.79$\pm$0.14 & 8.98$\pm$0.19 & 8.18$\pm$0.21 \\
 \ion{H}{II}-3 & 8.52$\pm$0.16 & 8.64$\pm$0.18 & 8.56$\pm$0.18 & 8.78$\pm$0.14 & 8.98$\pm$0.19 & 8.20$\pm$0.20 \\
 \ion{H}{II}-4 & 8.49$\pm$0.16 & 8.58$\pm$0.18 & 8.45$\pm$0.18 & 8.60$\pm$0.14 & 8.80$\pm$0.19 & 8.33$\pm$0.20 \\
 \ion{H}{II}-5 & 8.48$\pm$0.16 & 8.56$\pm$0.18 & 8.41$\pm$0.18 & 8.55$\pm$0.14 & 8.76$\pm$0.19 & 8.46$\pm$0.20 \\
 \ion{H}{II}-6 & 8.49$\pm$0.16 & 8.58$\pm$0.18 & 8.42$\pm$0.18 & 8.56$\pm$0.14 & 8.79$\pm$0.19 & 8.52$\pm$0.20 \\
 \ion{H}{II}-7 & 8.50$\pm$0.16 & 8.60$\pm$0.18 & 8.41$\pm$0.18 & 8.55$\pm$0.14 & 8.76$\pm$0.19 & 8.51$\pm$0.20 \\
\hline
\end{tabular}
\caption{The metallicity obtained with different calibrations of the N2 and the O3N2 indicators. For each indicator we list the value obtained using the \cite{Marino2013} method in columns 2 and 4 (M13), the  \cite{PettiniPagel2004} method in columns 3 and 5 (PP04), the $R_{23}$ method \citep{Pagel1979} for the upper limit in column 6 and the $S_{23}$ indicator \citep{PerezMontero2005} in the last column.}\label{tab:App2}
\end{table*}

\begin{table*}
    \centering
    \begin{tabular}{lcccccc}
    \hline
      Region &           12 + log(O/H) &          12 + log(Ne/H) &           12 + log(S/H) &          12 + log(Ar/H) &           12 + log(N/H)  &                    He/H  \\
      \hline
 \ion{H}{II}-1 &                      -- &                      -- &                      -- &                      -- &                      --  &                      --  \\
 \ion{H}{II}-2 &                      -- &                      -- &                      -- &                      -- &                      --  &                      --  \\
 \ion{H}{II}-3 &                      -- &                      -- &                      -- &                      -- &                      --  &                      --  \\
 \ion{H}{II}-4 &     8.25 $\pm$ 0.29     &     7.49 $\pm$ 0.74     &                      -- &                      -- &                      --  &                      --  \\
 \ion{H}{II}-5 &     8.10 $\pm$ 0.14     &     7.11 $\pm$ 0.30     &     6.91 $\pm$ 0.09     &     6.10 $\pm$ 0.15     &     7.61 $\pm$ 0.18      &   0.0809 $\pm$ 0.0146    \\
 \ion{H}{II}-6 &     8.01 $\pm$ 0.11     &     6.99 $\pm$ 0.22     &     7.15 $\pm$ 0.13     &     6.32 $\pm$ 0.18     &                      --  &   0.0849 $\pm$ 0.0227    \\
 \ion{H}{II}-7 &     8.01 $\pm$ 0.10     &     7.05 $\pm$ 0.24     &     6.96 $\pm$ 0.13     &     6.11 $\pm$ 0.24     &                      --  &   0.0706 $\pm$ 0.0228    \\
\hline
\end{tabular}
\caption{Elemental abundances obtained with the `direct' method (see text).}\label{tab:App3}
\end{table*}

\end{appendix}

\end{document}